\documentclass[pdflatex,sn-mathphys]{sn-jnl}
\usepackage{xcolor}
\usepackage{pdflscape}

\newcommand{\Lya}{Ly$\rm{\alpha}$}

\newcommand{\fesc}{$f_{esc}(\rm{Ly\alpha})$ }

\newcommand{\Hb}{H$\rm{\beta}$ }
\newcommand{\Ha}{H$\rm{\alpha}$ }
\newcommand{\OIII}{[OIII]$_{4959,5007}$ }

\newcommand{\simname}{{\it Azahar}}

\theoremstyle{thmstyleone}%
\theoremstyle{thmstyletwo}%

\theoremstyle{thmstylethree}%

\usepackage[symbol]{footmisc}

\usepackage[superscript, biblabel, nomove]{cite}
\usepackage{caption}

\begin{document}

\title[Deciphering Lyman-$\alpha$ Emission Deep into the Epoch of Reionisation]{Deciphering Lyman-$\alpha$ Emission Deep into the Epoch of Reionisation}

\author*[1,2]{\fnm{Callum} \sur{Witten}}\email{cw795@cam.ac.uk}
\author[2,3]{\fnm{Nicolas} \sur{Laporte}}
\author[4]{\fnm{Sergio} \sur{Martin-Alvarez}}
\author[1,2]{\fnm{Debora} \sur{Sijacki}}
\author[1,2]{\fnm{Yuxuan} \sur{Yuan}}
\author[1,2]{\fnm{Martin G.} \sur{Haehnelt}}
\author[2,3]{\fnm{William M.} \sur{Baker}}
\author[5]{\fnm{James S.} \sur{Dunlop}}
\author[6]{\fnm{Richard S.} \sur{Ellis}}
\author[7]{\fnm{Norman A.} \sur{Grogin}}
\author[8]{\fnm{Garth} \sur{Illingworth}}
\author[9]{\fnm{Harley} \sur{Katz}}
\author[7]{\fnm{Anton M.} \sur{Koekemoer}}
\author[10]{\fnm{Daniel} \sur{Magee}}
\author[2,3,6]{\fnm{Roberto} \sur{Maiolino}}
\author[2,3]{\fnm{William} \sur{McClymont}}
\author[11]{\fnm{Pablo G.} \sur{P\'erez-Gonz\'alez}}
\author[2,3]{\fnm{D\'{a}vid} \sur{Pusk\'{a}s}}
\author[12]{\fnm{Guido} \sur{Roberts-Borsani}}
\author[13]{\fnm{Paola} \sur{Santini}}
\author[2,3]{\fnm{Charlotte} \sur{Simmonds}}

\affil[\small 1]{\small \orgdiv{Institute of Astronomy}, \orgname{University of Cambridge}, \orgaddress{\street{Madingley Road}, \city{Cambridge}, \postcode{CB3 0HA}, \country{UK}}} 

\affil[\small 2]{\small \orgdiv{Kavli Institute for Cosmology}, \orgname{University of Cambridge}, \orgaddress{\street{Madingley Road}, \city{Cambridge}, \postcode{CB3 0HA}, \country{UK}}}

\affil[\small 3]{\small \orgdiv{Cavendish Labratory}, \orgname{University of Cambridge}, \orgaddress{\street{Madingley Road}, \city{Cambridge}, \postcode{CB3 9BB}, \country{UK}}}

\affil[\small 4]{\small \orgdiv{Kavli Institute for Particle Astrophysics and Cosmology}, \orgname{Stanford University}, \orgaddress{\city{Stanford}, \postcode{CA 94305}, \country{USA}}}

\affil[\small 5]{\small \orgdiv{Institute for Astronomy}, \orgname{University of Edinburgh, Royal Observatory}, \orgaddress{\city{Edinburgh}, \postcode{EH9 3HJ}, \country{UK}}}

\affil[\small 6]{\small \orgdiv{Department of Physics and Astronomy}, \orgname{University College London}, \orgaddress{\street{Gower Street}, \city{London}, \postcode{WC1E 6BT}, \country{UK}}}

\affil[\small 7]{\small \orgdiv{Space Telescope Science Institute}, \orgaddress{\street{3700 San Martin Drive}, \city{Baltimore}, \postcode{MD 21218}, \country{USA}}}

\affil[\small 8]{\small \orgdiv{Department of Astronomy and Astrophysics}, \orgname{University of California}, \orgaddress{\city{Santa Cruz}, \postcode{CA 95064}, \country{USA}}}

\affil[\small 9]{\small \orgdiv{Department of Physics}, \orgname{University of Oxford}, \orgaddress{\street{Denys Wilkinson Building, Keble Road}, \city{Oxford}, \postcode{OX1 3RH}, \country{UK}}}

\affil[\small 10]{\small \orgdiv{UCO/Lick Observatory}, \orgname{University of California}, \orgaddress{\city{Santa Cruz}, \postcode{CA 95064}, \country{USA}}}

\affil[\small 11]{\small \orgdiv{Centro de Astrobiolog\'{\i}a}, \orgname{CSIC-INTA}, \orgaddress{\street{Ctra. de Ajalvir km 4, Torrej\'on de Ardoz}, \city{Madrid}, \postcode{E-28850}, \country{Spain}}}

\affil[\small 12]{\small \orgdiv{Department of Physics and Astronomy}, \orgname{University of California}, \orgaddress{\city{Los Angeles}, \postcode{CA 90095}, \country{USA}}}

\affil[\small 13]{\small \orgdiv{INAF - Osservatorio Astronomico di Roma}, \orgaddress{\street{via di Frascati 33, 00078 Monte Porzio Catone}, \country{Italy}}}

\abstract{ \bf
During the epoch of reionisation the first galaxies were enshrouded in pristine neutral gas, with one of the brightest emission lines in star-forming galaxies, Lyman-$\alpha$ (\Lya), expected to remain undetected until the Universe became ionised. Providing an explanation for the surprising detection of \Lya\ in these early galaxies is a major challenge for extra-galactic studies. Recent JWST observations have reignited the debate on whether residence in an overdensity of galaxies is a {\it sufficient and necessary} condition for \Lya\ to escape. Here, we take unique advantage of both high-resolution and high-sensitivity images from the \textit{JWST} instrument NIRCam to reveal that all galaxies in a sample of $z>7$ \Lya\ emitters have {\it close companions}. We exploit novel on-the-fly radiative transfer magnetohydrodynamical simulations with cosmic ray feedback to show that galaxies with frequent mergers have very bursty star formation which drives episodes of high intrinsic \Lya\ emission and facilitates the escape of \Lya\ photons along channels cleared of neutral gas. We conclude that the rapid build up of stellar mass through mergers presents a compelling solution to the long-standing puzzle of the detection of \Lya\ emission deep into the epoch of reionisation.
}

\maketitle

Young, vigorously star-forming galaxies have been identified in the very early Universe \cite{Curtis-Lake_2023, Arrabal-Haro23, Williams23}. These galaxies should be excellent sources of Lyman-$\alpha$ emission (\Lya, $\lambda$=1215.67$\text{\AA}$) -- the intrinsically brightest emission line \cite{2014PASA...31...40D} -- which stems from the recombination of hydrogen that has been ionised by their young stellar populations. However, deep in the epoch of reionisation galaxies are expected to be exceptionally gas-rich such that their stellar nurseries are enshrouded in copious amounts of neutral hydrogen, which leads to extreme damped absorption of \Lya\ \cite{Heintz+23}. Furthermore, the intergalactic medium (IGM) is increasingly neutral as we probe to higher redshift \cite{Robertson_2015, deBarros_2017} and this neutral gas is expected to resonantly scatter \Lya\ emission. Hence, due to `local' attenuation by a gas-rich interstellar medium (ISM) and scattering by a neutral IGM, \Lya\ emission should only be detectable towards the end of the reionisation era, about one billion years after the Big Bang \cite{2014PASA...31...40D, Endsley22, Bolan22, Hsiao23}. 

While the decreasing observability of \Lya\ emission with increasing redshift has been repeatedly claimed to be observed \cite{Stark+17, deBarros+2019,Jones+23}, this picture has been challenged by the occasional, surprising detection of \Lya\ emission in several galaxies deep in the reionisation era \cite{Bunker+23, Jung2023}. It has been suggested that \Lya\ can escape through the neutral IGM if the galaxies reside in sufficiently large ionised bubbles embedded within the neutral IGM, driven either by active galactic nuclei (AGN) \cite{Laporte+17,Larson23,Maiolino+23_AGN} or by an enhanced radiation field produced by an overdensity of associated objects \cite{Castellano+16, Tilvi+20, Leonova+22, Morishita+22, Tang+23, Tacchella+23, Witstok+23}. However, a solution to the escape of \Lya\ emission through the ISM and CGM of what are expected to be very gas-rich galaxies remains elusive. Before the advent of \textit{JWST}, the sensitivity and resolution of imaging instruments meant that studies of \Lya\ emitters (LAEs) at high redshift were spatially unresolved. Hence, it was not possible to probe the physical processes that could explain the escape of \Lya\ emission from the ISM.

To address this crucial issue we study a sample of nine galaxies that have been spectroscopically confirmed with the detection of \Lya~emission at $z>7$ with high-resolution spectrographs and which have been observed with publicly-available \textit{JWST} Near Infrared Camera (NIRCam) \cite{Rieke+23} imaging (the properties of which are reported in Table~\ref{tab:Galaxy_Properties}). These galaxies fall within the GOODS-North, GOODS-South, EGS and COSMOS fields, and have been observed as part of five different programs: PRIMER (PI: Dunlop), FRESCO (PI: Oesch)\cite{Oesch+23}, CEERS (PI: Finkelstein)\cite{Bagley+23}, JADES (PI: Eisenstein)\cite{Rieke+23b} and DDT program 4426 (providing NIRSpec IFU observations of GN-z11, PI: Maiolino)\cite{Maiolino+23_PopIII}. Six of these nine \Lya-emitting galaxies are known to lie within overdensities \cite{Tang+23,Jung+2022,Leonova+22,Tacchella+23,Witstok+23}, three of which also likely host accreting black holes \cite{Laporte+17,Larson23,Maiolino+23_AGN}. Moreover, it has been shown that the remaining galaxies are incapable of alone blowing a large enough ionised bubble to facilitate the escape of \Lya\ through the neutral IGM, suggesting that an underlying population of faint galaxies must be surrounding these LAEs \cite{Witstok+23}.

However, the presence of spectroscopically confirmed galaxies within known ionised bubbles that do not show \Lya\ emission \cite{Tang+23, Jones+23, Witstok+23} indicates that there {\it must} be further {\it local} processes at play driving \Lya\ emission deep into the epoch of reionisation. With this in mind, we use the unparalleled sensitivity and resolution of \textit{JWST}/NIRCam in order to accurately determine the properties of these nine LAEs, with equivalent width (EW) greater than 10\AA, at $z > 7$.

 \captionsetup[figure]{labelfont={bf},name={Figure},labelsep=period}
 \setcounter{figure}{0}    

\begin{figure}
    \includegraphics[width=1.\columnwidth]{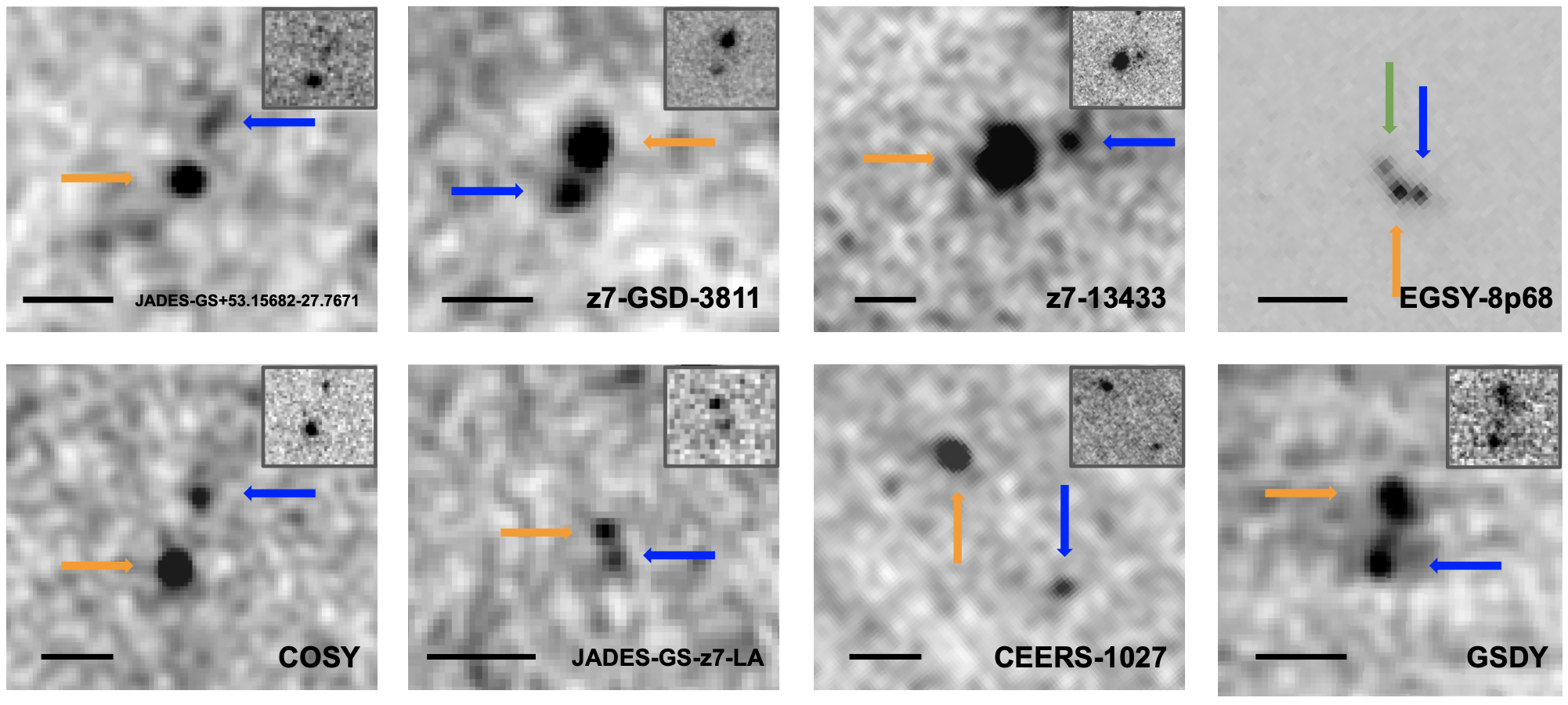}\\
    \caption{{\bf An abundance of galaxy mergers seen with NIRCam.} Cutouts of \textit{JWST} images (as different surveys make use of different filters, the filter within which the companion is most clearly present is shown: F182M for GSDY and z7-GSD-3811, F200W for EGSY8p68, F115W for CEERS-1027, JADES-GS-z7-LA and COSY and F150W for JADES-GS+53.15682-27.7671 and z7-13433) showing the multiple components of each system. All images are smoothed with a Gaussian of FWHM $\sim 0.7$ kpc to enhance the visibility of the companion, except for EGSY8p68 where the three components become unresolved if we smooth the images. We include the unsmoothed images as sub-panels in the top right of each panel in order to emphasise that the companion is always visible before smoothing. We use different colorbar ranges for each system in order to make the companion galaxy well defined, hence some unsmoothed images can have what appears like a nosier background than others. The position of the main target (*A in the text) is displayed by an orange arrow, the position of the first companion (*B) is indicated by a blue arrow and the third companion, if any, by a green arrow. The name of each candidate is indicated in the bottom right corner, and in the bottom left of each panel is a black scale showing 0.5$^{\prime \prime}$.}
    \label{fig:1}
\end{figure}

Remarkably, NIRCam images of these galaxies reveal the presence of multiple components to all of these LAEs, as shown in Figure~\ref{fig:1}. Given the tight redshift constraints provided by strong \OIII emission falling into medium band filters, we estimate the probability of these being unrelated, high-redshift objects and find this is always below $14 \%$ for each of our systems and is often just $1 \%$ (see Methods for further details). We estimate the redshift of each candidate companion by carefully extracting the Spectral Energy Distribution (SED) of both the main component and its companions in Extended Figure~\ref{Efig:1} and ~\ref{Efig:2}. We then fit each SED by assuming several parametric star formation histories  (SFH - burst, constant, delayed, exponential), allowing a large range for all parameters \cite{Carnall2018}. In all cases, we deduce the photometric redshifts of the companions to be consistent with that of their spectroscopically-confirmed main component.

\begin{landscape}
\begin{table}
    \centering
    \begin{tabular}{ l | ccccccc }
    \hline
    \hline
ID 	&	z	&	log($\rm{M_{*}[M_{\odot}]}$)	&   SFR $\rm{[M_{\odot}/yr]}$	& Separation [kpc] & L$_{Ly\alpha}$ [10$^{43}$ erg/s] & \fesc & Reference \\ 
\hline
COSY-A & \textit{7.142} & 8.77 $^{+ 0.02 }_{- 0.02 }$ & 5.91 $^{+ 0.27 }_{- 0.25 }$ &$-$ & 1.37 & $< 0.1$ &\cite{RB+16},\cite{Witten+23} \\
COSY-B$^\dagger$ & \textit{7.142} & 7.59 $^{+ 0.13 }_{- 0.11 }$ & 0.39 $^{+ 0.14 }_{- 0.08 }$ &2.8 & $-$ & $-$ \\ 
JADES-GS-z7-LA-A & \textit{7.278} &   6.72 $^{+ 0.14 }_{- 0.12}$ & 1.13$^{+0.05}_{-0.03}$ & $-$ & 0.15 & $0.96 \pm 0.22$ &\cite{Saxena+23}\\
JADES-GS-z7-LA-B & 7.27$^{+0.07}_{-0.06}$  &  7.68$^{+0.18}_{-0.19}$  & 2.44$^{+0.82}_{-0.47}$ & 1.2 &$-$&$-$&\\
z7-13433-A & \textit{7.482} & 9.62 $^{+ 0.01 }_{- 0.01 }$ & 42.19 $^{+ 1.09 }_{- 1.00 }$  & $-$ & 1.00 & $-$ & \cite{Jung+2022} \\
z7-13433-B & \textit{7.478} & 8.67 $^{+ 0.03 }_{- 0.04 }$ & 4.65 $^{+ 0.33 }_{- 0.36 }$ & 2.6 &$-$& $-$ \\
z7-GSD-3811-A & \textit{7.661} & 8.96 $^{+ 0.12 }_{- 0.13 }$ & 9.25 $^{+ 2.79 }_{- 2.39 }$ &$-$ & 0.386 &$ 0.22 \pm 0.08$ &  \cite{Song+2016}, this work \\
z7-GSD-3811-B & \textit{7.658} & 8.77 $^{+ 0.21 }_{- 0.21 }$ & 5.88 $^{+ 3.73 }_{- 2.28 }$ & 1.0 & $-$ & $-$ \\
CEERS-1027-A & \textit{7.819} & 7.94 $^{+ 0.05 }_{- 0.04 }$ & 0.87 $^{+ 0.10 }_{- 0.08 }$  & $-$ &0.432& $0.085 \pm 0.018$ & \cite{Tang+23} \\
CEERS-1027-B & 8.07 $^{+ 0.12 }_{- 0.14 }$ & 7.90 $^{+ 0.28 }_{- 0.21 }$ & 0.80 $^{+ 0.74 }_{- 0.30 }$ & 8.6 &$-$& $-$ \\
GSDY-A & \textit{7.957} & 8.75 $^{+ 0.21 }_{- 0.27 }$ & 17.80 $^{+ 4.56 }_{- 5.02 }$& $-$ & $0.19$ &  $>0.11$ & \cite{RB+22}, this work \\
GSDY-B & \textit{7.958} &7.75 $^{+ 0.21 }_{- 0.28 }$ &  2.73 $^{+ 1.02 }_{- 0.73 }$ & 2.2 &$-$& $-$&  \\
JADES-GS-A$^\ddagger$ & \textit{7.982} & 7.82$^{+0.11}_{-0.09}$ & 3.97$^{+1.78}_{-0.88}$& $-$ & 0.056 & $0.09 \pm 0.01$ & \cite{Jones+23}\\
JADES-GS-B$^\ddagger$ & 8.11$^{+0.10}_{-0.08}$  & 8.33$^{+0.14}_{-0.16}$ & 10.47$^{+5.16}_{-3.46}$& 1.5 & $-$& $-$&\\
EGSY8p68-A & \textit{8.683} & 8.88 $^{+ 0.10 }_{- 0.18 }$ & 7.62 $^{+ 1.87 }_{- 2.54 }$ & $-$& 1.59 & $< 0.1$ & \cite{Zitrin+2015},\cite{Witten+23}\\
EGSY8p68-B$^\dagger$ & 8.40$^{+0.37}_{-0.27}$ & 8.34 $^{+ 0.28 }_{- 0.36 }$ & 2.22 $^{+ 1.99 }_{- 1.25 }$ & 0.5 & $-$ & $-$ & \\
EGSY8p68-C$^\dagger$ & 8.74$^{+ 0.50 }_{- 0.46 }$ & 8.63 $^{+ 0.26 }_{- 0.40 }$ & 4.26 $^{+ 3.49 }_{- 2.55 }$ & 0.6 &$-$ & $-$ &\\
GN-z11-A & \textit{10.603} & 9.1 $^{+ 0.3 }_{- 0.4 }$ & 21 $^{+ 22 }_{- 10 }$ &$-$&0.324& 0.03 $^{+ 0.05 }_{- 0.02 }$ & \cite{Bunker+23} \\
GN-z11-B & \textit{10.62} & * & * & 1.6 & $-$ & $-$ & \cite{Scholtz+23}\\
GN-z11-C & \textit{10.603} & * & * & 2.5 & $-$ & $-$ & \cite{Maiolino+23_PopIII}\\

 \hline
    \end{tabular}
    \caption{{\bf The properties of \Lya-emitting galaxies.} Physical properties of all the systems studied in this paper. The columns are : (1) spectroscopic (italic) or photometric redshift of the candidate, (2) stellar mass, (3) Star formation rate averaged over at most the last 100 Myr, (4) separation between the main component (A) and each companion, (5) luminosity of observed \Lya\ emission (6) escape fraction of Ly$\alpha$ photons and (7) reference for the original spectroscopic confirmation and the Ly$\alpha$ escape fraction measurement. * The companions of GN-z11 are spectroscopically confirmed \cite{Scholtz+23, Maiolino+23_PopIII} however given that they are very UV-faint we cannot estimate their properties. $^\dagger$ The companions of both COSY and EGSY8p68 have recently been spectroscopically confirmed by NIRSpec IFU observations \cite{Ubler+23};[Carniani et al. in prep.]. $^\ddagger$ JADES-GS+53.15682-27.76716 has been abbreviated to JADES-GS.}
    \label{tab:Galaxy_Properties}
\end{table}
\end{landscape}

To spectroscopically confirm their identity as companions we exploit existing XSHOOTER/VLT \cite{Vernet2011}, MOSFIRE/Keck \cite{MOSFIRE1,MOSFIRE2}, Near-Infrared Spectrograph (NIRSpec) on-board \textit{JWST} \cite{Jakobsen22} and NIRCam wide-field slitless spectroscopy (WFSS) observations \cite{Rieke2005} of these galaxies. For three systems in our sample we have no spectroscopic information on their companions given the limitations of seeing at ground-based telescopes and the small aperture size of JWST/NIRSpec observations. For the remaining six systems in our sample we are able to spectroscopically confirm the companion galaxy as it falls within the FoV of the observations of the main LAE. For two companions in our sample we see tentative ($\sim 3\sigma$) evidence of \Lya\ emission in Extended Figure~\ref{Efig:3} (one of which has recently been confirmed by NIRSpec IFU observations \cite{Ubler+23}. For a further two companions we observe the [OIII]$_{5007}$ emission line at $>8\sigma$ (as seen in Extended Figure~\ref{Efig:4} with fluxes reported in Extended Table~\ref{tab:ELfluxes}), and NIRSpec IFU observations spectroscopically confirm the companions of the final two systems \cite{Maiolino+23_PopIII, Scholtz+23} and [Carniani et al. in prep.]. We therefore conclude that for all systems for which we have resolved spectroscopic observations of the companion ($\sim67\%$), we spectroscopically confirm its nature as a physical companion (see the Methods for further details of the spectrosocopic analysis). 

To confirm that the presence of a close companion is the primary factor governing the visibility of \Lya, we determine the fraction of companions seen in a mass-matched sample of $z>7$ galaxies with high-resolution spectroscopic data for which \Lya\ is not detected \cite{Jones+23}. We find that 43\% of these galaxies have photometric-candidate companions within 5$^{\prime \prime}$ of the central galaxy which is consistent with the companion fraction determined by a more comprehensive study [Pusk\'{a}s et al., in prep]. The lower fraction of close companions among samples that are not selected for \Lya\ emission is evidence that the 100\% rate of companions for our sample of LAEs is atypical of $z>7$ galaxies.

To reinforce the observational evidence supporting the idea that ongoing interactions drive an increase in the detectability of \Lya\ emission during the epoch of reionisation, we explore comparable galaxy mergers using the novel \simname~simulation suite (Martin-Alvarez et al., in prep.). This simulation suite was performed before we obtained the NIRCam data, but remarkably, we find many simulated galaxies that match the photometry and the spatial geometry of our objects. \simname~is a cosmological, high-resolution, zoom-in simulation which employs a magnetohydrodynamical solver together with state-of-the-art cosmic ray feedback (see the pathfinder Pandora project \cite{Martin-Alvarez+2023} and Methods).  Most importantly for this work, \simname~also features on-the-fly radiative transfer \cite{Rosdahl2015, Rosdahl2018, Rosdahl2022}, capable of self-consistently reproducing reionisation while fully modelling the ISM of galaxies (with a maximum spatial resolution of $10$~pc), and crucially resolving the propagation and escape of ionising radiation on the ISM scales \cite{Kimm2014, Garel2021}. We post-process our simulations with the publicly available RASCAS code \cite{Michel-Dansac20} to account for both the production and resonant scattering of \Lya\ photons as well as scattering or absorption of \Lya\ photons by dust \cite{Laursen09b}.

\begin{figure}
    \centering
    \includegraphics[width=\columnwidth]{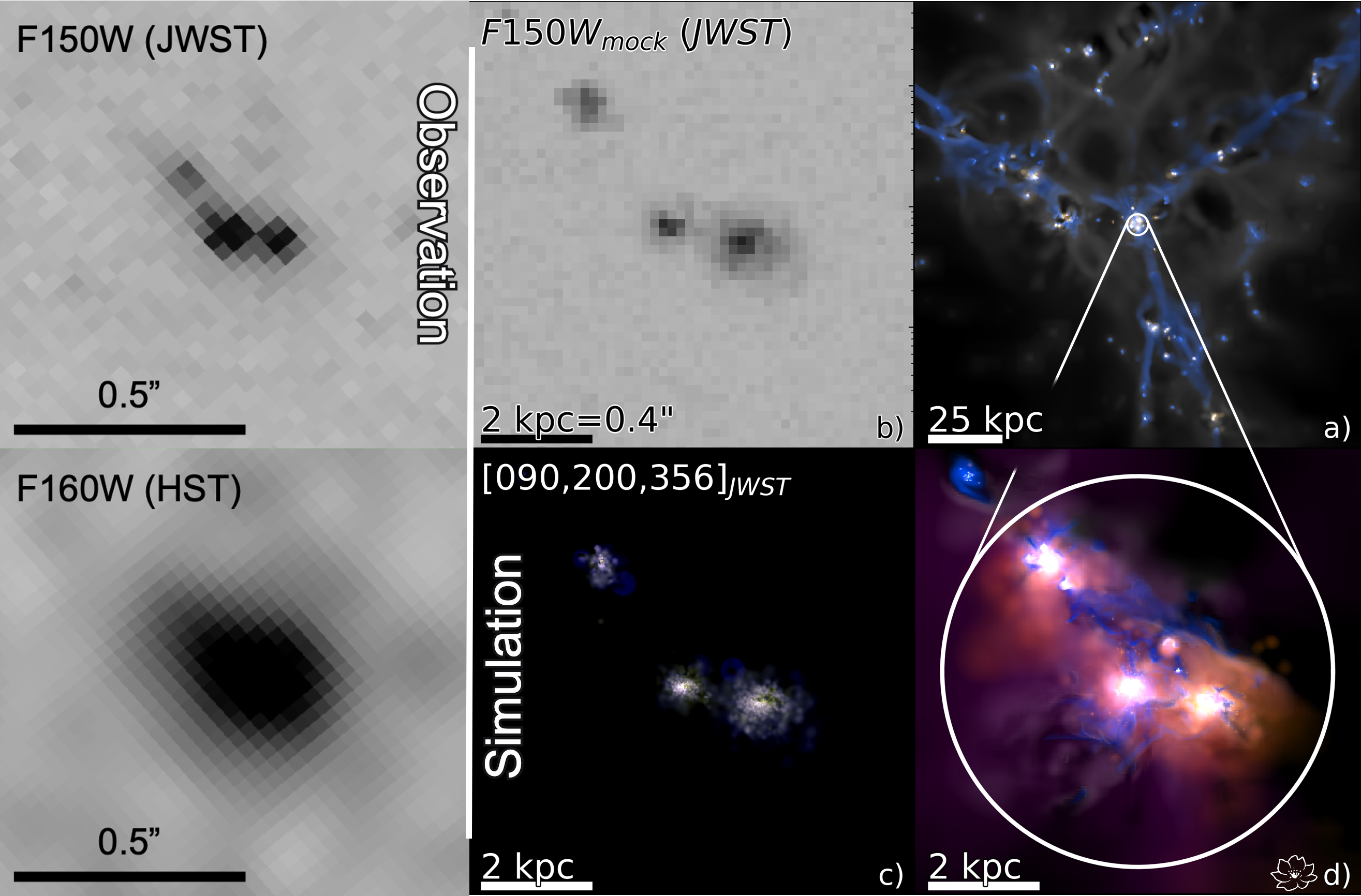}\\
    \caption{{\bf Comparison of \simname~to an observed galaxy merger.} {\bf (Left panels)} NIRCam F150W (above) and \textit{HST} F160W (below) imaging of the LAE EGSY8p68. The NIRCam imaging reveals three components to the system that were previously unresolved by \textit{HST}. {\bf (Right panels)} Analogue galaxy merger from the \simname~simulation: {\it a)} large-scale view of the filaments encompassing the galaxy merger (blue: HI density, grey-black: HII density, yellow-white: F150W intensity); {\it b)} simulated NIRCam F150W observation of the \simname~merger; {\it c)} fully resolved simulated observation in the NIRCam filters; and {\it d)} hydrogen gas and ionising radiation properties (blue: HI density, grey-black: HII density, yellow-white: F150W intensity, purple: LyC radiation energy density, orange: HeII ionising radiation energy density). The appearance of the observed system is very well reproduced by an ongoing merger of three galaxies, with an extended escaping LyC radiation field.}
    \label{fig:2}
\end{figure}

\begin{figure}
    \centering
    \includegraphics[width=0.95\columnwidth]{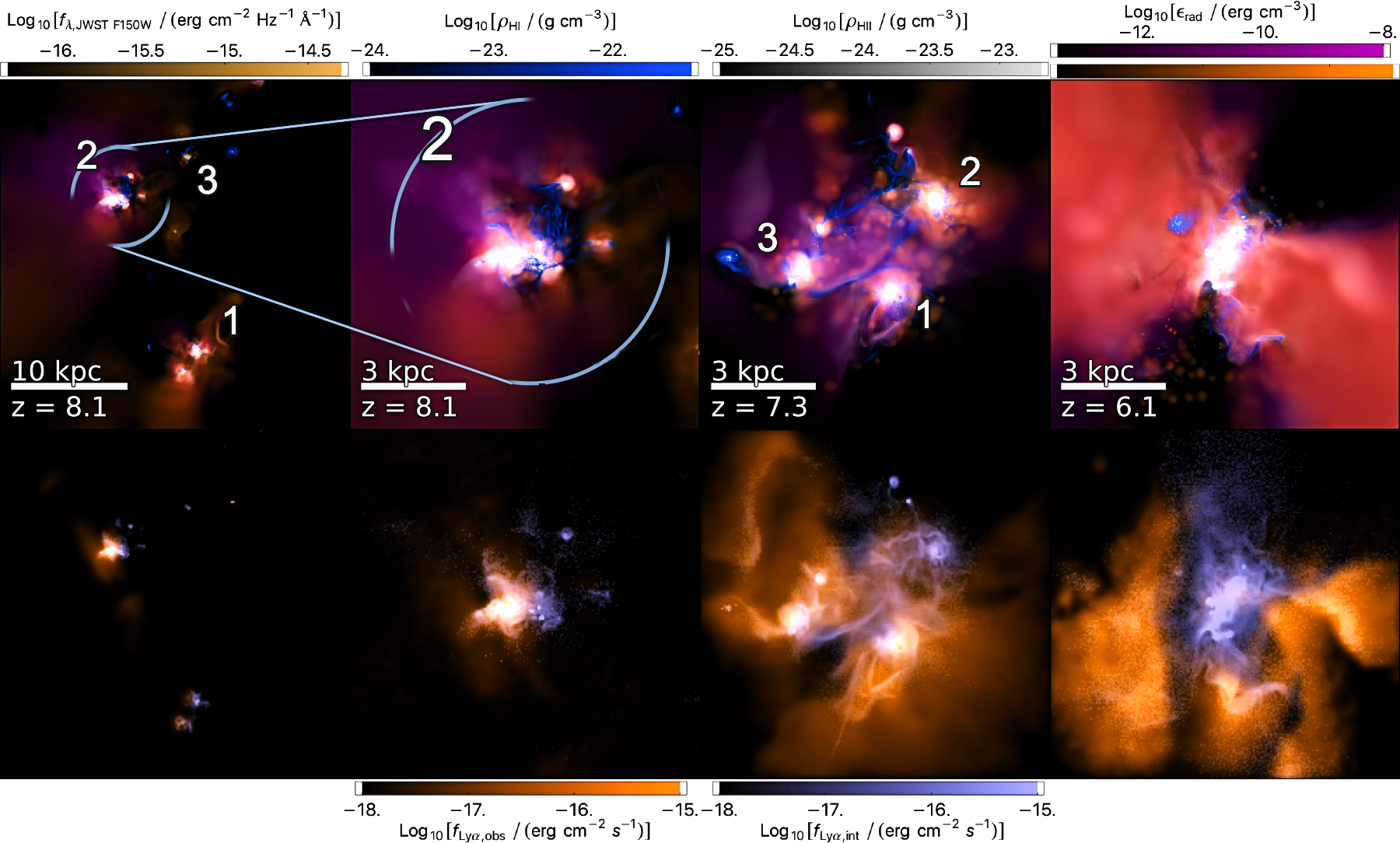}\\
    \includegraphics[width=1.0\columnwidth]{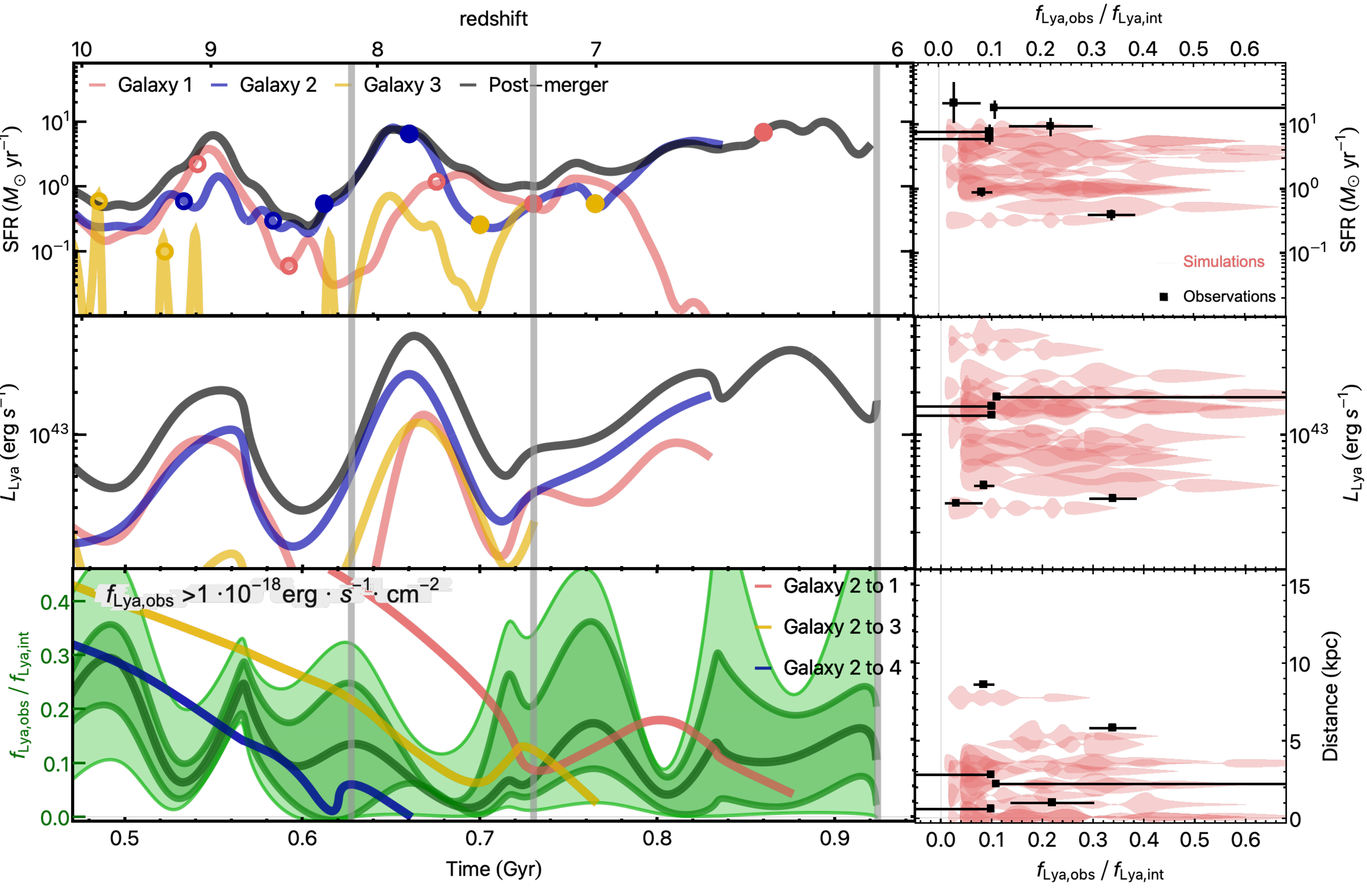}\\
    \caption{{\bf The evolution of galaxy properties with an ongoing merger.} {\bf (Rows 1-2)} The time evolution of our simulated merger system at $z = 8.1$ (close up and large scale view), $7.3$ and $6.1$. First row: maps of HI density (blue), HII density (grey-black), NIRCam F150W intensity (yellow-white), LyC radiation energy density (purple) and HeII ionising radiation energy density (orange). Second row: the RASCAS intrinsic \Lya\ flux (blue) and scattered \Lya\ flux (orange). {\bf (Rows 3-5)} The SFR for each of the three galaxies and the final system (black line). Dot symbols show mergers with companions (filled markers) and with other secondary systems (open markers); the intrinsic \Lya\ luminosity of the three galaxies and the entire system; the fraction of escaped \Lya\ luminosity filtered for pixels with fluxes higher than a given limit (median (dark green curve), quartiles (darker band), and min-max of the distribution across 12 lines-of-sight (lighter band)). Distance between the main progenitor and its companions is shown as well. The grey vertical lines in Rows 3-5 correspond to the redshifts shown in Rows 1-2. The rightmost panels show key physical relations for the observations (black) and the simulations (pink).}
    \label{fig:3}
\end{figure}

Due to a substantial overdensity of galaxies in our simulated volume, the main progenitor undergoes repeated mergers with multiple other galaxies brought in by the cosmic filaments, often involving several companions or mergers in rapid succession as has been observed in \cite{Asada+23}. To highlight one such merger occurring at $z \sim 7.3$, the four rightmost panels of Figure~\ref{fig:2} show different projections of \simname. This particular interaction features three merging galaxies that will constitute the main progenitor of the spiral galaxy formed by $z \sim 1$. These are very good analogues to the EGSY8p68 observations shown on the left (modelling  of the observations with \textsc{Galfit} in Extended Figure~\ref{Efig:5} confirms the presence of three components), with very similar individual galaxy sizes, mutual distances and merger configuration. Specifically, the three simulated galaxies have stellar masses of $M_{*,1} = 2 \cdot 10^{8}\,M_\odot$, $M_{*,2} = 3 \cdot 10^{8}\,M_\odot$, and $M_{*,3} = 8 \cdot 10^{7}\,M_\odot$ at this redshift, which are comparable to the stellar masses of the EGSY8p68 galaxies (see Table~\ref{tab:Galaxy_Properties}). Furthermore, the simulated SFRs vary between $1-10\,\rm{M_\odot yr^{-1}}$ (shown in Figure~\ref{fig:3}) and are consistent with the SFRs of the EGSY8p68 galaxies ($7.62\, \rm{M_\odot yr^{-1}}, 2.22\, \rm{M_\odot yr^{-1}}, 4.26\, \rm{M_\odot yr^{-1}}$). We also consider the total \Lya\, luminosity of the simulated system and find that, while it varies, its average value is $\sim 10^{43}$~erg/s (shown in Figure~\ref{fig:3}), in agreement with the value for EGSY8p68, reported in Table~\ref{tab:Galaxy_Properties}. Finally we note that simulated volume hosts a substantial overdensity of galaxies, which is consistent with the local environments of the observed LAEs in our sample (as discussed earlier), including EGSY8p68 \cite{Leonova+22}. We thus conclude that the simulated system matches all key observed photometric and spectrosopic properties of EGSY8p68 (see also Methods).

The observations shown on the left clearly indicate the superior resolution and sensitivity of \textit{JWST} (top panel) over \textit{HST} (bottom panel). The tri-component nature of EGSY8p68 is entirely unresolved in existing \textit{HST} observations, but clearly identifiable in the F150W NIRCam imaging. Panel (\textit{a}) shows the large-scale view (150 kpc across) of the three filaments encompassing the merging system, where large amounts of HI gas (blue) along the filaments are feeding the star formation in these systems, resulting in significant NIRCam F150W emission (yellow-white). This active star formation is driving ionised hydrogen bubbles (grey) away from the filaments and into the low density regions. Panel (\textit{b}) shows the simulated \textit{JWST}-like 150W observation (with dust extinction modelled as an absorption screen along the line-of-sight), where we select the line-of-sight to illustrate the resemblance with EGSY8p68. Note that the simulated merging system is displayed at a slightly earlier stage of the merger with respect to EGSY8p68, with our simulated galaxies approaching the likely physical separation of the observed galaxies at $z \sim 7$. Interestingly, the mock F150W emission reveals compact galactic cores analogous to these \textit{JWST} observations, with diffuse emission (blending with the background) emerging from the extended, low density stellar discs. Panel (\textit{c}) shows a synthetic colour-composite observation using the \textit{JWST} filters at the full resolution of our simulation. While the two companion galaxies are more diffuse, with stars actively forming in their discs, the main galaxy has a much more compact core due to the preceding rapid growth through repeated mergers. Panel (\textit{d}) is a colour composite of various simulated properties relevant for \Lya\ detectability. The emission from LyC ionising photons (purple) and HeII ionising photons (orange) is distributed on a scale of a few (physical) kpc. Importantly, there is a clear separation of the stellar emission and the HI gas during the merger event, with the ionising radiation escaping from star forming regions. 

To provide a quantitative confirmation and in-depth physical understanding of this effect, we explore in Figure~\ref{fig:3} the time evolution of the three merging galaxies in detail. In the top row the colours encode the same quantities as in panel (\textit{d}) of Figure~\ref{fig:2}, while in the second row we show the intrinsic \Lya\ emission (blue) and resonantly scattered \Lya\ emission (orange), produced by post-processing our simulations with RASCAS. The first column shows a larger scale view of the main progenitor (galaxy 2) and its approaching companions at $z \sim 8.1$. The second column shows a close up views of the main progenitor at $z \sim 8.1$ undergoing a minor merger, with a complex topology of neutral gas and channels through which \Lya\ photons are escaping. The third column shows the ongoing merger of our galaxies 1, 2 and 3 at $z \sim 7.3$ (with a different orientation with respect to Figure~\ref{fig:2}), where bursty star formation drives ionised channels through the ISM, with a significant amount of gas also tidally stripped from the galaxies. The fourth column shows a post-merger view of the newly formed disc-dominated galaxy at $z \sim 6.1$. Here, the HI is confined to the merger remnant galaxy, and the radiation can only escape through the channels opened by collimated, cosmic ray-driven bi-conical outflows. Focusing now on the \Lya\ radiation, all three galaxies have significant intrinsic \Lya\ emission due to their ongoing star formation. However, at this particular viewing angle, due to absorption by dust, \Lya\ scattered emission is only present around merging galaxy 2 at $z \sim 8.1$ and around galaxies 1 and 3 at $z \sim 7.3$. At $z \sim 7.3$, the remaining emission, not absorbed by dust, is very diffuse on large scales and hence not observable. 

To understand what drives the observable \Lya\ emission, in the third row we show the SFR evolution for our three merging galaxies, together with the SFR history of the merger remnant at $z \sim 6.1$. SFRs of all three galaxies undergo repeated cycles of bursts, driven by numerous mergers (`major' mergers [filled dots] and `minor' mergers [empty dots]) and a high SFR `plateau' during the final merger episode from $z \sim 7$ to $6$. This is mirrored in the evolution of the intrinsic \Lya\ emission of each galaxy as well as the total emission integrated across our three simulated galaxies, as shown in the fourth row, where there is a clear correspondence between the SFR peaks and peaks of intrinsic \Lya. {\it Our simulations hence demonstrate that high SFR triggered by gas-rich mergers results in brighter intrinsic \Lya\ emission, which enhances the probability of these systems being detected.} 

To constrain this more robustly, we select an observable \Lya\ flux threshold of $10^{-18} \rm{erg}\,\rm{s}^{-1}\,\rm{cm}^{-2}$ consistent with our observations, and compute the ratio of scattered to intrinsic \Lya\ emission over the spatial extent of our mock observations of the galaxies. This is shown in the fifth row, where the dark green curve denotes the median of the ratio of scattered to intrinsic \Lya\ flux and the shaded bands (quartiles and minimum-maximum values) show its distribution computed over 12 different lines-of-sight. As the fraction of escaping \Lya\ emission has a number of high peaks we conclude that the \Lya\ emission should be detectable for a number of the redshifts studied here. Interestingly, during close galactic encounters (as shown by the distance trajectories of three main galaxies shown in the same panel by coloured lines) and mergers, {\it significant gas tidal stripping and bursty star formation feedback leads to the opening of more low-column density sight-lines, hence facilitating the escape of \Lya\ radiation}. This effect can be clearly seen when comparing the most important merger instances to the peaks (high-tails) in the distribution of \Lya\ escape fraction. Our findings regarding higher intrinsic \Lya\ emission and enhanced \Lya\ escape fraction in mergers are robust to the assumed dust model (see Methods and Extended Figure~\ref{Efig:6}). It is worth noting that inclusion of AGN feedback in our simulations would only reinforce our findings, as expected black hole fuelling during mergers would power AGN \cite{Costa_2022}.

The rightmost panels of Figure~\ref{fig:3} show various physical relations for the observations (black points) and the simulations (pink `violin' symbols, denoting the distribution of \Lya\ escape fraction over 12 lines-of-sight), using the same flux filter as above, demonstrating a very good correspondence between our simulations and our observed LAEs. Interestingly, both for observations and simulations we do not find any clear correlations between the \Lya\ escape fraction and any other galaxy properties (distance, SFR, stellar mass), further corroborating our finding that `favourable' lines-of-sights with evacuated neutral hydrogen channels lead to directional \Lya\ photon leakage. 

In this paper, we introduce a new interpretation to explain the unexpected detection of \Lya\ in $z\geq 7$ galaxies in an epoch where the IGM is mostly neutral and gas-rich, early galaxies are heavily `locally' \Lya\ damped. Combining the high-resolution and high-sensitivity of \textit{JWST} data with state-of-the-art radiative transfer on-the-fly magnetohydrodynamics simulations, we demonstrate that {\it three ingredients} are key to making \Lya\ emission detectable from our sample of galaxies deep in the epoch of reionisation: \textit{galactic mergers} driving high intrinsic \Lya\ emission in the host galaxy; \textit{a `favourable' line-of-sight} cleared of local neutral hydrogen in the host galaxy by tidal interactions with companions and by star-formation feedback; a sufficiently large ionised bubble facilitating the escape of \Lya\ emission through the IGM.

\newpage

\noindent{\bf \large Methods}

 \captionsetup[figure]{labelfont={bf},name={Extended Figure},labelsep=period}
 \setcounter{figure}{0}    

 \captionsetup[table]{labelfont={bf},name={Extended Table},labelsep=period}
 \setcounter{table}{0}    

\vspace{0.2cm}
\noindent{\bf \large Data} 

\noindent The emergence of \textit{JWST} with its significant technical advancements over both ground- and space-based telescopes, has the potential to reveal new hints as to the driving forces behind the observed \Lya~emission discussed in this paper. We therefore decided to analyse all spectroscopically-confirmed LAEs at $z > 7$, with \textit{JWST}/NIRCam imaging. This search originally revealed eleven such LAEs, however, further constraints on the SED of these galaxies from NIRCam imaging reveals the \Lya~emission line detected in two of these LAEs is in fact not \Lya~(discussed in more detail below). Therefore, the sample becomes nine LAEs (seen in Table~\ref{tab:Galaxy_Properties}), eight of which have companions that are identifiable in publicly available NIRCam imaging and GN-z11 that has spectrosocpically confirmed companions that do not show a UV continuum \cite{Maiolino+23_PopIII, Scholtz+23}. These galaxies are found in the GOODS-North and GOODS-South, EGS and COSMOS fields which have been observed with NIRCam by FRESCO (PI: Oesch, ID: 1895) and JADES (PI: Eisenstein, ID: 1180), CEERS (PI: Finkelstein, ID: 1345) and PRIMER (PI: Dunlop, ID: 1837), respectively.

The PRIMER survey covers $\sim 150 \rm{arcmin}^{2}$ of the COSMOS field with the F090W, F115W, F150W, F200W, F277W, F356W, F410M and F444W filters all at a $5 \sigma$ limiting depth of $m_{\rm AB} > 28$. The CEERS survey took imaging of $100 \rm{arcmin}^{2}$ of the EGS field in the F115W, F150W, F200W, F277W, F356W, F410M and F444W filters, reaching a $5 \sigma$ limiting depth of $m_{\rm AB} > 29$ in the short-wavelength or $m_{\rm AB} > 28.4$ in the long-wavelength filters \cite{Finkelstein+22}. The June 2023 JADES data release surveys a $\sim 36$ arcmin$^2$ area of the GOODS-S field using the F090W, F115W, F150W, F200W, F277W, F335M, F356W, F410M and F444W, reaching $5 \sigma$ limiting depths of $m_{\rm AB} \sim 30$ \cite{Eisenstein+23}.

The FRESCO survey's primary aim is to obtain NIRCam Long-Wavelength (LW) Grism spectra over the F444W filter for galaxies across GOODS-S and GOODS-N. However, the simultaneous imaging in the short-wavelength channel provides imaging in both F182M and F210M filters, and direct imaging also obtained in the F444W filter additionally provides us with a total of 3 filters for our galaxies in GOODS-S/GOODS-N. FRESCO's LW WFSS brings significant advantages as the F444W filter covers the wavelength range of $3.9-5 \mu$m, therefore observing both \OIII and \Hb at $z \sim 7-9$. These emission lines ordinarily contaminate imaging in the F444W filter and hence constraints on the fluxes of these emission lines allow us to reduce the uncertainty in galaxy properties such as stellar mass, star-formation rate and age. We use these constraints on the emission lines present in F444W to create a mock F430M filter. Furthermore, when we fit the SED (discussed further below) we check the fluxes of the \OIII and \Hb emission lines in the best-fit SED model and confirm that these are consistent with the measured fluxes from the WFSS spectrum. 

\vspace{2cm}
\noindent{\bf Data Reduction}

\noindent Throughout our analysis we use the final reduced data products created by the PRIMER team for the COSMOS field, the publicly available reduced data products from the CEERS team for the EGS field, the publicly available reduced data products from the JADES team for the deep HST region of GOODS-S and our own reduction of the GOODS-S FRESCO imaging for regions outside of the publicly available JADES imaging. In order to produce the GOODS-S images, we take all available NIRCam imaging data of GOODS-S, available from the FRESCO survey, and reduce them using version 1.8.5 of the \textit{JWST} pipeline. We first download the uncalibrated files from the MAST archive and follow the steps of the \textit{JWST} pipeline: (Stage 1) detector-level corrections and ramp fitting, (Stage 2) instrument-level and observing-mode corrections producing fully calibrated exposures, (Stage 3) producing the final mosaic. We take additional care after Stage 1 to ensure the removal of horizontal and vertical striping that can be present in the rate files. 

We extract the photometry of our objects using \textsc{SEXtractor} \cite{SEXtractor} on PSF-matched images extracting all objects with more than 4 pixels (DETECT\_MINAREA 4) detected at more than 1.5$\sigma$ (DETECT\_THRESH and ANALYSIS\_THRESH 1.5). We used large deblending parameters to allow efficient separation of each component (DEBLEND\_NTHRESH 64 and DEBLEND\_MINCOUNT 0.000001), however, the small separation between the components of the EGSY8p68 and JADES-GS-z7-LA systems makes the extraction of each component's photometry impossible with \textsc{SEXtractor}. For JADES-GS-z7-LA we employ the photometry of the LAE and its companion reported in \cite{Saxena+23} while for EGSY8p68 we model the shape of each component using \textsc{Galfit} \cite{Galfit} assuming a Sersic profile and NIRCam PSFs, simulated using \textsc{WebbPSF}. We first run \textsc{Galfit} on one of the NIRCam images where all the components are clearly resolved (F115W), and we then fix the resultant position, Sersic index, axis ratio and angle for each component when extracting the photometry for the remaining images. Extended Figure~\ref{Efig:5} shows the modelling of the components for this system.

\begin{figure}
    \centering
    \includegraphics[width=1.0\columnwidth]{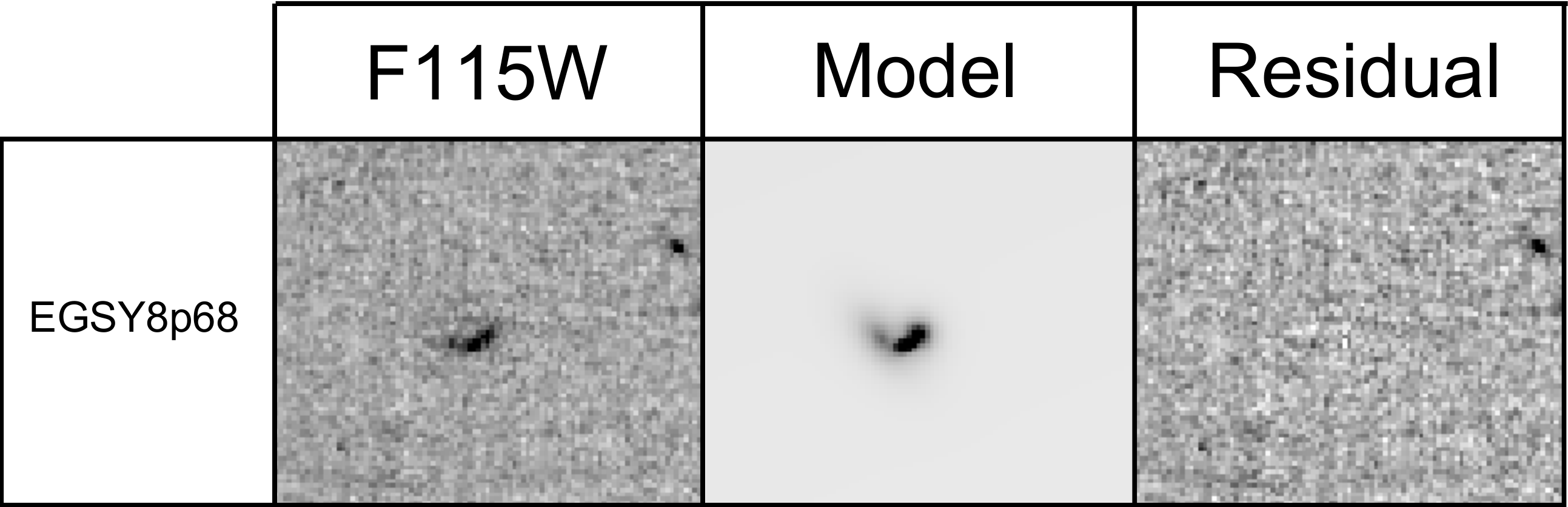}%
    \caption{Example of \textsc{Galfit} modelling for the EGSY8p68 system for which the separation between each component is too small for the photometry to be extracted by \textsc{SEXtractor}. }
    \label{Efig:5}
\end{figure}

As discussed earlier, it is desirable to spectroscopically confirm as many companion galaxies as possible. The use of the Mutli-Shutter Array with NIRSpec to observe three of the systems (EGSY8p68, CEERS-1027 and JADES-GS+53.15682-27.76716) means that, given the small shutter size, the companion is not coincident with the shutter position. Instead, we make use of all existing ground-based observations of our sample of LAEs, of which four of the sample have been observed by MOSFIRE (z7-13433 - ID: N199, PI: Jung; EGSY8p68 - ID: C228M, PI: Zitrin; GSDY - ID: U069, PI: Treu; z7-GSD-3811 - ID: C182M, PI: Scoville) and one with X-shooter (COSY - ID: 097.A-0043, PI: Ellis). 

We reduce the spectroscopic data using the standard MOSFIRE data reduction pipeline and EsoReflex. These pipelines perform wavelength callibration, sky subtraction and flat fielding as well as further standard data reduction steps in order to produce two-dimensional (2D) spectra. The resultant 2D spectra span the length of the original slit used for the observations and have spectral and spatial resolutions of $R \sim 3380$, 0.1798 $^{\prime \prime}$/px and $R \sim 8900$, 0.158 $^{\prime \prime}$/px for MOSFIRE and X-shooter respectively, but are limited by the seeing of each observation which can be as poor as $\sim 0.9$$^{\prime \prime}$.

Two of our sample are located in the GOODS-S field (z7-GSD-3811 and GSDY). This field was the target of WFSS observations by the NIRCam Long-wavelength (LW) Grism as part of the FRESCO observing program. FRESCO's LW WFSS covers the wavelength range of $3.9-5 \mu$m using the F444W filter, therefore observing both the \OIII and \Hb emission lines for $z \sim 7-9$ galaxies, which are expected to be bright in high-redshift galaxies. These lines have previously been observed in $z > 7$ galaxies using the publicly available FRESCO data \cite{Laporte+23, Oesch+23}. Moreover, the significantly superior spatial resolution of NIRCam LW WFSS (60 milli-arcesconds) allows us to resolve emission lines where MOSFIRE and X-shooter would fail to do so. 

We reduced the Grism data following the steps described in \cite{Sun+22} including flat fielding, background subtraction, 1/f noise subtraction and WCS assignment. We also perform a careful astrometric analysis to avoid any offsets between the LW direct imaging in F444W and the LW channel Grism spectra in F444W. This allows us to extract the 2D Grism spectra associated with the astrometric position of any source, these spectra are then stacked and the 1D spectrum at the position of the source with an aperture of 3 pixels is extracted. We also perform a continuum subtraction from the 1D spectrum to remove any contaminant continuum and we focus on measuring \OIII and \Hb emission line fluxes. 

\begin{figure}
    \hspace{-1cm} \includegraphics[width=1.2\columnwidth]{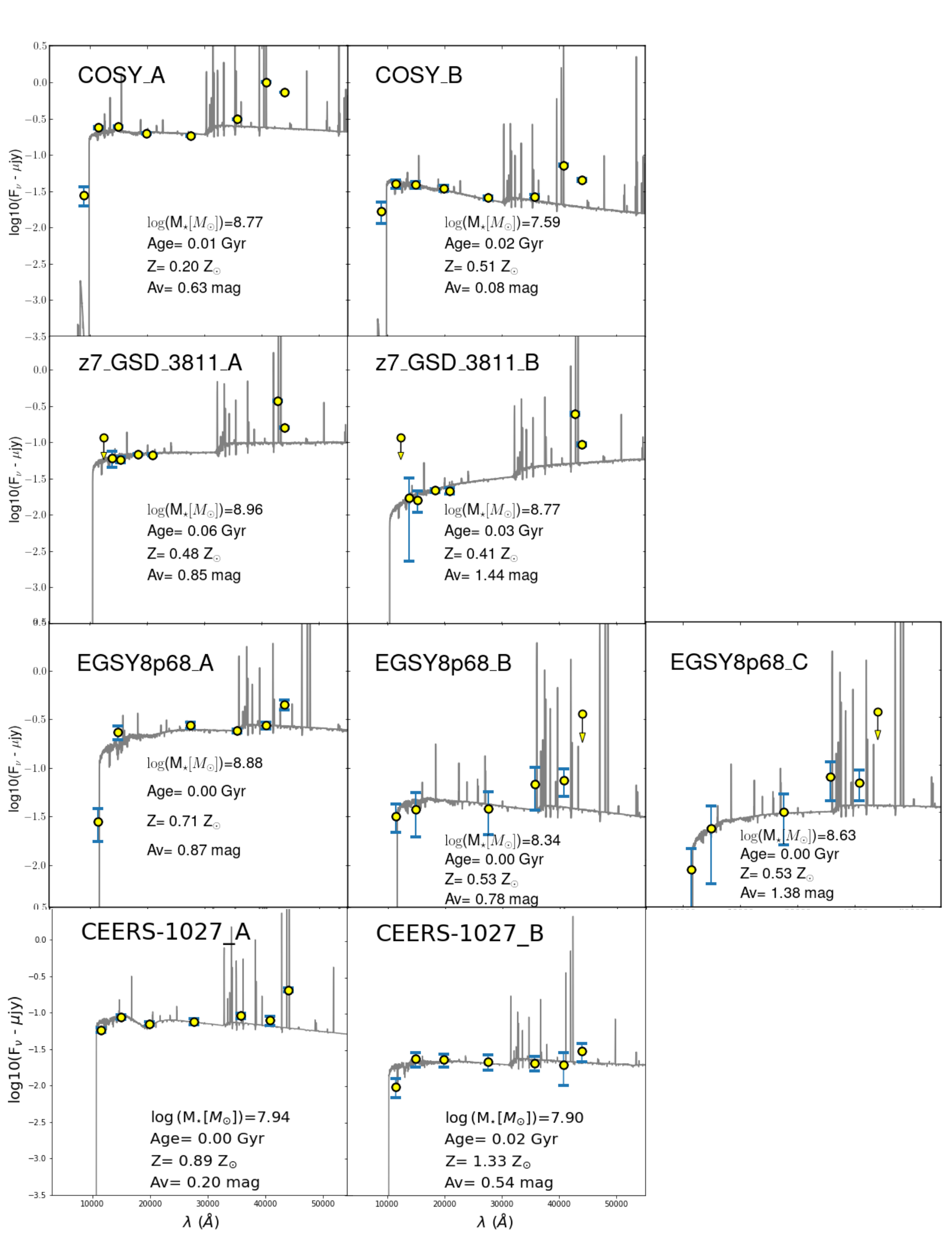} \\
    \caption{Best SED-fits for all systems with strong Ly$\alpha$. The yellow dots are the observed photometry, the grey line shows the best fit. The physical parameters of the best fit are also indicated (see text for details).}
    \label{Efig:1}
\end{figure}

\begin{figure}
     \hspace{-1cm}  \includegraphics[width=0.8\columnwidth]{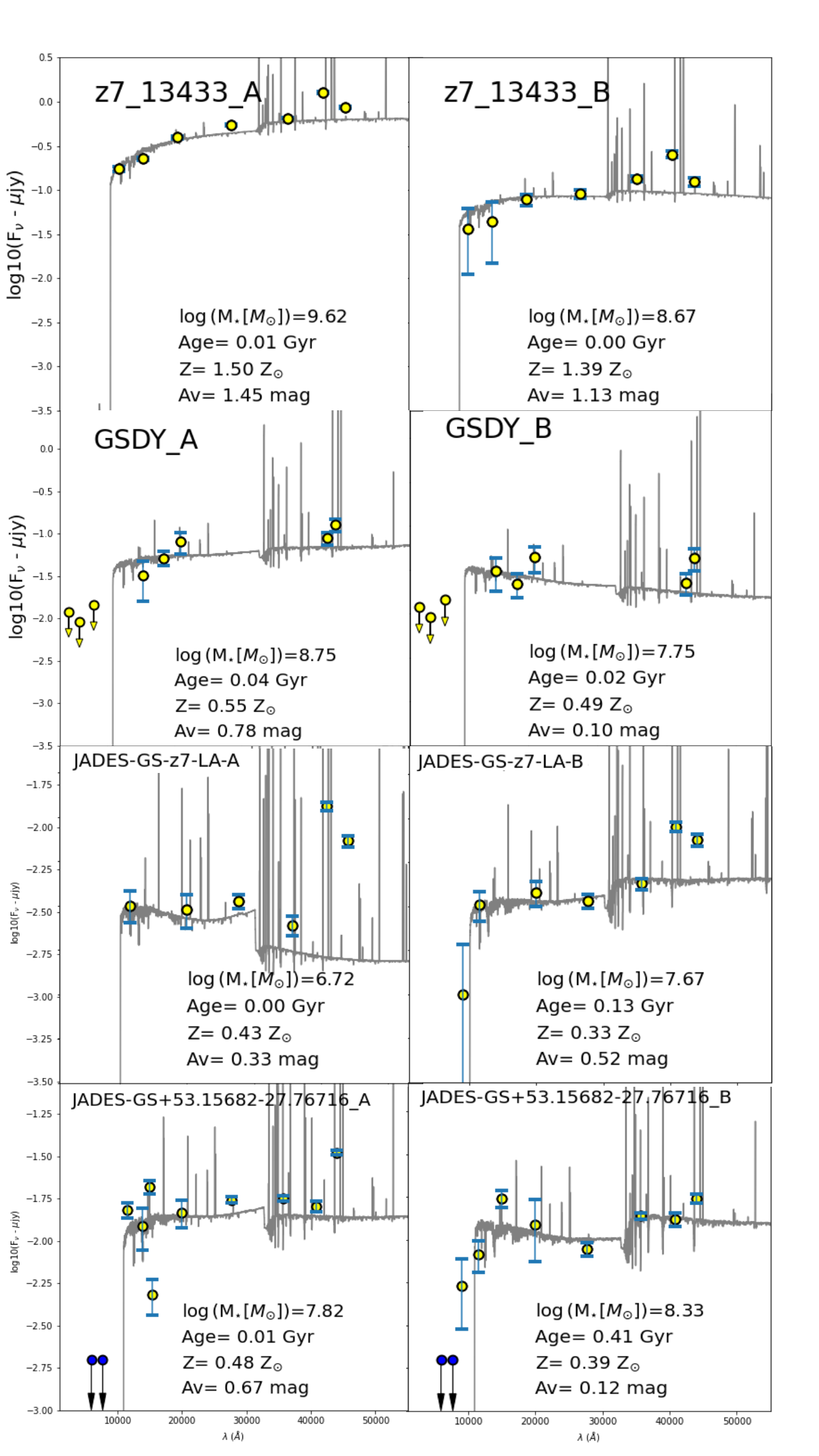} \\
    \caption{Same as Extended Figure~\ref{Efig:1}.}
    \label{Efig:2}
\end{figure}

\vspace{0.2cm}
\noindent{\bf Redshift Determination} 

\noindent We initially take care to ensure that, for galaxies that only have \Lya~emission detected in their spectra, the emission line detected is indeed \Lya. There are initially five LAEs at $z>7$ in the literature, imaged by NIRCam, for which \Lya\ is the only observed emission line (z7-GSD-3811, z8-19326, z7-13433, z7-20237 and GSDY). In order to confirm that this emission line is \Lya, we use the SED-fitting code \textsc{Bagpipes} \cite{Carnall2018} on the available NIRCam data to produce an improved photometric redshift of each LAE. One of the key diagnostics in this fitting becomes the bright \OIII emission lines, expected from high redshift objects, that fall into either the medium-band filter F410M (for $z\sim 6.8-7.6$)/F430M (for $z\sim 7.3-7.8$) or a lower flux consistent with continuum emission in this filter but a boost in the F444W filter (for $z\sim 6.8-9$). This photometric-redshift fitting results in a redshift that is consistent with the redshift of the emission lines observed being \Lya\ in all but two of these galaxies (z8-19326 and z7-20237). The photometric redshift, constrained by a significant boost in F410M ($z \sim 7.1$) for z8-19326, becomes inconsistent with the emission line detected being \Lya~and therefore this galaxy is not included in our sample. While boosts in both F356W and F444W strongly indicate that z7-20237 is in fact at $z \sim 6$. The original HST photometry reported in \cite{Jung+2022} also slightly favoured a $z \sim 6$ solution making the observed emission line likely to be C IV. The removal of these two galaxies as contaminants underlines the challenges in determining galaxy redshifts with just one emission line and wide-band HST filters pre-JWST. The \textit{JWST} offers significant advancement in this area given NIRSpec's sensitivity allowing the detection of multiple lines, but also given the frequency of programs that exploit the NIRCam medium-band filters around $\sim 4\mu$m. All of the galaxies in our final sample show a clear \OIII excess in NIRCam imaging. We can therefore be confident that the emission line detected is indeed \Lya.

Following this previous step, in order to confirm that all components of each system are at a similar redshift, we again employ \textsc{Bagpipes} to estimate their photometric redshifts. We then compare these with the spectroscopic redshift of the main component. This fitting confirms that all the companion galaxies preferentially have high photometric redshifts, all of which are consistent with the spectroscopic redshift of the massive galaxy. In combination with the prior of the companion being closely located to a confirmed high-redshift galaxy, these become strong-photometric candidate companion galaxies. We also exploit all available ground and space based spectroscopic observations that cover the companions and use these to spectroscopically confirm the companion galaxy where possible. Below we discuss the available redshift constraints on our sample.

\textbf{EGSY8p68, CEERS-1027 and JADES-GS+53.15682-27.76716} constitute the sub-sample of LAEs that have photometric candidate companions with no spectroscopic confirmation. This lack of confirmation is due to a combination of factors. The observations of CEERS-1027 and JADES-GS+53.15682-27.76716 were made with the NIRSpec MSA and the positioning of the shutter misses the companion, and as such there is no ability to place any spectroscopic constraints on these companions. EGSY8p68 has been observed by both the NIRSpec MSA and the ground-based telescope MOSFIRE/Keck. The MOSFIRE observations of the system cover the companion galaxies, however due to the seeing of the observations ($\sim 0.7$$^{\prime \prime}$) and the small separation between the components of the system ($\sim 0.1$$^{\prime \prime}$) resolving the components was not possible. Follow-up observations with the NIRSpec MSA reveal EGSY8p68 is likely an AGN given the presence of high-ionisation emission lines and a significant broadening of the \Hb emission line \cite{Larson23}. While one of the companion galaxies lies in the shutter of this observation, the available spatial resolution and strength of the emission lines from EGSY8p68 makes it challenging to disentangle the emission from each component. However, the reported NIRSpec MSA flux is seven times fainter than that observed in the MOSFIRE spectrum \cite{Zitrin+2015}. Given the MSA shutter size is significantly smaller than the MOSFIRE slit, one solution to this discrepancy in flux is that some of the \Lya\ emission is extended outside of the MSA slit. We note that the component EGSY8p68-C sits outside of the MSA slit position and as such this component may additionally host \Lya\ emission at the same redshift as the main component. Moreover, \cite{Larson23} claim that the system is consistent with being involved in a major merger and recent NIRSpec IFU observations of EGSY8p68 spectroscopically confirm they are physical companions [Carniani et al. in prep.]. However, given these results are not yet published and the data is not yet public we continue our analysis of the redshift constraints assuming we do not have this spectroscopic confirmation.

Despite the lack of spectroscopic constraints on this sub-sample, the similarities in the SEDs of the main and companion galaxies are distinct. The galaxies in each system show similar colours, Lyman-breaks that are consistent and an emission-line boost, driven by \OIII and \Hb emission, in the same filters, as can be seen in Extended Figure~\ref{Efig:1} and~\ref{Efig:2}. This results in a photometric redshift for each of the companions that clearly favours a $z > 7$ solution, and given the emission line feature in the F444W filter, the photometric redshift is further constrained to $z = 7.6 - 9$. 
While the redshift of these components are very confidently constrained to within this $\Delta z = 1.4$, further constraining their redshift is challenging without spectroscopy. However, we make use of the UV luminosity function from \cite{Bouwens+15} in order to estimate the expected number of galaxies within the volume that the \OIII emission redshift constraint constrains the companion galaxies to. We note that the photometric redshift from BAGPIPES (reported in Table~\ref{tab:Galaxy_Properties}) is much more tightly constrained than that provided by the \OIII boost alone, however we use this wider redshift constraint ($\Delta z = 1.4$) to provide an upper limit on the expected number of galaxies. Given that we have searched for companions in a 3$^{\prime \prime}$x3$^{\prime \prime}$ region, we use this as the area for this estimate and the \OIII redshift constraints to produce a volume. We then integrate the $z=7$ or $z=8$ (dependent on the spectroscopic redshift of the main component) UV luminosity function down to the 5$\sigma$ magnitude limit of the NIRCam images. We estimate that the expected number of galaxies in this field-of-view that are consistent with \Hb and \OIII producing the excess F444W flux and with observed UV magnitude is $0.09^{+0.16}_{-0.04}$ for JADES-GS+53.15682-27.76716, $0.007^{+0.009}_{-0.002}$ for CEERS-1027 and $0.003^{+0.006}_{-0.001}$ for EGSY8p68. It is therefore very unlikely that another galaxy that is not associated with the spectroscopically confirmed LAE would be present in this volume.

\textbf{JADES-GS-z7-LA} is an extremely high EW LAE (400 \rm{\AA}) discovered in JADES NIRSpec observations of galaxies in the GOODS-S field \cite{Saxena+23}. The presence of a photometric candidate companion galaxy to this LAE has already been reported \cite{Saxena+23}, but the companion lies outside of the NIRSpec MSA shutter, and hence is not spectroscopically confirmed. We perform the same analysis as above, noting that the \OIII emission is in the F410M filter, limiting the redshift to between $z = 6.8 - 7.6$, and find the expected number of galaxies to be $0.13^{+0.14}_{-0.05}$. Therefore, once again, it is unlikely that the observed companion is at a redshift that makes it unrelated to the spectroscopically confirmed LAE. Moreover, we note the observed excess in the F115W filter that may be driven by \Lya\ emission in both the main and companion galaxy \cite{Saxena+23}, supporting the conclusion that these two galaxies are at the same redshift.

\textbf{COSY and z7-13433} have companions that show excess fluxes in the F410M filter driven by \OIII emission. SED-fitting of their companions strongly prefers a $z > 7$ solution, and the presence of \OIII emission confines the redshift to $z = 6.8 - 7.6$. We again perform the same analysis as above, estimating the expected number of galaxies in the FoV. The expected number of galaxies is $0.03^{+0.03}_{-0.01}$ for COSY and $0.04^{+0.03}_{-0.01}$ for z7-13433 and thus we conclude it is very likely that the central and companion galaxies are associated with each other. 

Moreover, the companions of both COSY and z7-13433 are within the slits used for the observations with X-shooter and MOSFIRE, respectively. The two components of each system are shown in Extended Figure~\ref{fig:3}, where we can identify both the positive (white) and negative (black) traces of the objects, caused by the ABBA nodding pattern of the observations. We find that, for z7-13433, two emission lines are offset by $\sim 3$ pixels, which is consistent with the $\sim 0.5$$^{\prime \prime}$ offset between the two components in the slit. The \Lya~emission that is coincident with the position of the companion is detected at $2.8 \sigma$ and corresponds to $z = 7.479$, while the \Lya~flux associated with the main galaxy is at $4 \sigma$ corresponding to $z = 7.482$. Therefore, these two components are offset by $\sim 125$ km/s. In the case of COSY, we expect a similar spatial offset between the two components, but the $\sim 0.7$$^{\prime \prime}$ seeing during the observations makes distinguishing between the two components very challenging. However, there is a second, previously unidentified, component to the \Lya~emission of COSY at $\sim 0.9898 \mu$m. We attribute this $2.7 \sigma$ emission line to COSY-B and note a velocity offset of $\sim 280$ km/s between COSY-A and COSY-B. While these \Lya~detections are clearly tentative, given they are coincident with the expected position in the 2D spectra of strong-photometric candidate galaxies and given their proximity to confirmed LAEs this boosts their likelihood of being a true detection. 

Finally, we note that recent NIRSpec IFU observations of COSY spectroscopically confirm its companion galaxy \cite{Ubler+23}. However, as with EGSY8p68, given these results are not yet published and the data is not yet public we continue our analysis without this spectroscopic information.

\begin{figure}
    \centering
    \includegraphics[width=1.0\columnwidth]{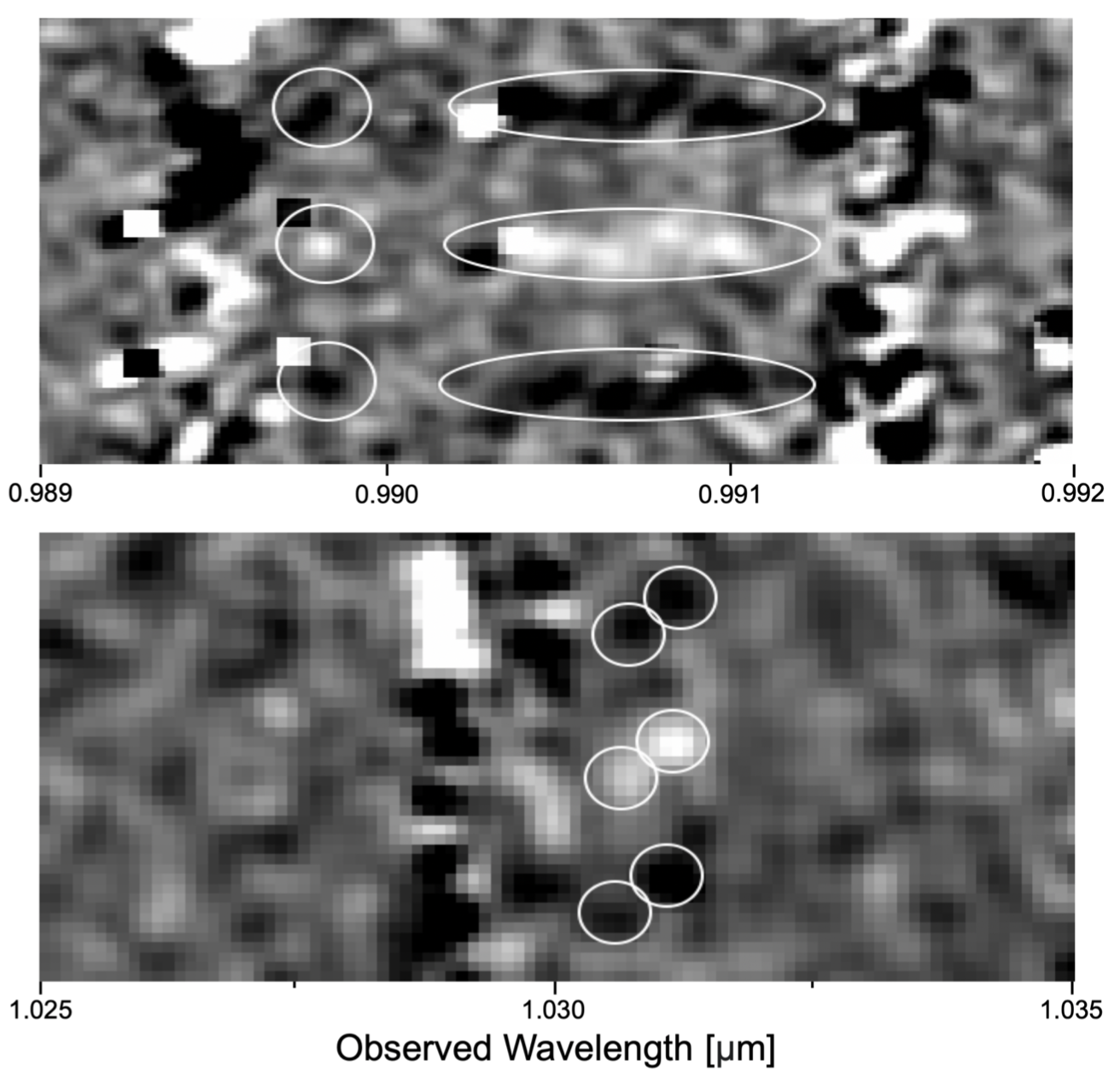}%
    \caption{The X-shooter (above) and MOSFIRE (below) Gaussian-smoothed 2D spectra of COSY and z7-13433, respectively. These cover the wavelength range of the previously spectroscopically confirmed \Lya\ emission of the system. White circular apertures indicate the proposed two components of the \Lya\ emission that are coincident with the expected positions of the main and companion galaxies in the slit. The positive trace of the emission is indicated by white pixels, the negative trace by black pixels, this effect of positive and negative traces is caused due to the ABBA nodding procedure employed by both instruments.}
    \label{Efig:3}
\end{figure}

\textbf{z7-GSD-3811 and GSDY} are both in the FoV of the FRESCO NIRCam WFSS survey. Their final 2D WFSS spectra, shown in Extended Figure~\ref{Efig:4} show \OIII emission lines for both z7-GSD-3811, GSDY and their companions. For z7-GSD-3811 both components of the \OIII emission are present and there is a $\sim 3 \sigma$, faint detection of H$\rm{\beta}$. Moreover, \OIII emission is also detected in the companion galaxy. This emission is offset from the central galaxy by $170 \pm 70$ km/s. The 2D spectrum of GSDY shows a clear detection of [OIII]$_{5007}$ in both the central and companion galaxies, and there is a very faint indication of [OIII]$_{4959}$ and \Hb in the companion galaxy, but at a less than 3$\sigma$ confidence for both. The emission detected in the central and companion galaxy are offset by just $\sim 26$ km/s, but given the low spectral resolution of the NIRCam WFSS, this is significantly smaller than the $\sim 70$ km/s uncertainty on this measurement. 

We discuss in the main text the use of \Hb to estimate the intrinsic \Lya\ flux and hence the escape fraction of \Lya\ photons, however, it can also be used as a probe of the instantaneous SFR. We assume a H$\alpha$ to \Hb ratio of 2.85 and Case B recombination at $T_{\rm e} = 10,000$ K and $N_{\rm{e}} = 10^4\ \rm{cm^{-3}}$ and following \cite{Kennicutt+98} we estimate the instantaneous SFR which is reported in Extended Table~\ref{tab:ELfluxes}. Given the lack of constraints on the redenning of these galaxies, we do not dust-correct the \Hb flux and as such these measured instantaneous SFRs are lower bounds. 

The observation of multiple, spatially resolved components to each emission line is taken as spectroscopic confirmation of the companion galaxies. Moreover, the velocity offset observed in all of these galaxies is indicative of either a line-of-sight separation, or more likely, a difference in the line-of-sight velocities of the two objects as they interact with each other. We therefore consider this as further evidence that the systems are indeed interacting, as opposed to being two components of the same stable system.

\begin{figure}
    \centering
    \includegraphics[width=1.0\columnwidth]{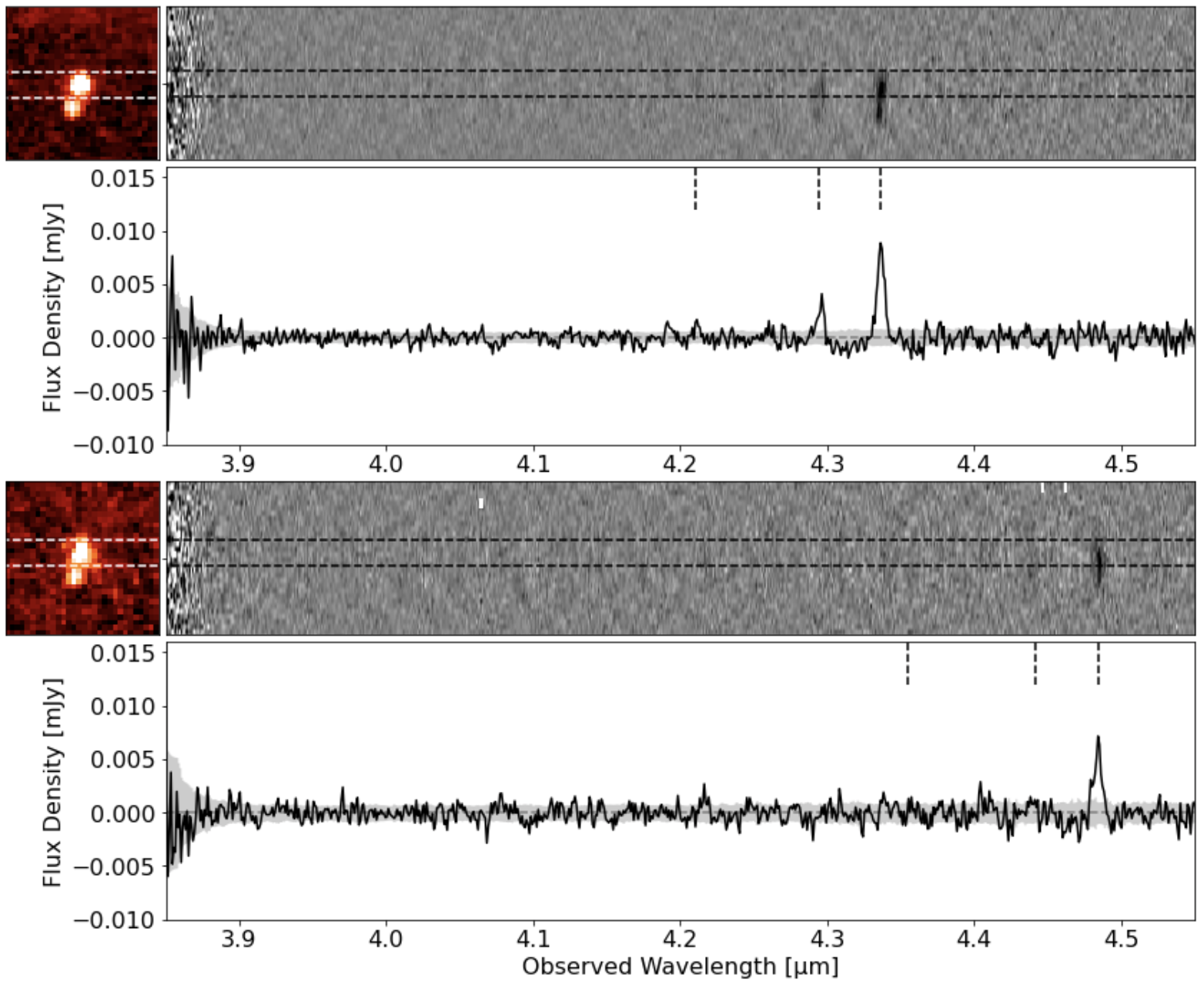}%
    \caption{NIRCam WFSS Grism spectra of z7-GSD-3811 (above) and GSDY (below). Within each panel, the continuum subtracted 2D spectrum (above) and 1D spectrum (below) of the central galaxy are shown. The upper left sub-panel shows the F444W image of the system where two components can be identified. The horizontal dashed lines indicate the position of the central galaxy in both the image and 2D spectrum. The three dashed lines in the 1D spectrum denote the expected positions of the \OIII doublet (central and right dashed lines) and \Hb (left dashed line) emission lines given the object's spectroscopic redshift, while the grey shaded region denotes the one-sigma noise level.}
    \label{Efig:4}
\end{figure}

\begin{table}[h]
    \centering
    \begin{tabular}{ | l | ccccc | }
    \hline
ID 	&	$z$	& $\rm{[OIII]_{5007}}$ & $\rm{[OIII]_{4959}}$ & \Hb & SFR$_{\rm{inst}} \rm{[M_{\odot}/yr]}$	\\  \hline
z7-GSD-3811-A & $7.663$ & $8.3 \pm 0.4$ & $2.7 \pm 0.4$ & $1.0 \pm 0.3$ & $16.5 \pm 4.9$\\
z7-GSD-3811-B & $7.658$ & $4.8 \pm 0.4$ & $1.5 \pm 0.3$ & $<0.5$ & $<8.2$\\
GSDY-A & $7.956$ & $5.8 \pm 0.7$ & $1.0 \pm 0.4$ & $<0.5$ & $<9.0$\\
GSDY-B & $7.957$ & $3.1 \pm 0.6$ & $1.5 \pm 0.6$ & $0.8 \pm 0.3$ & $14.4 \pm 5.4$\\

 \hline
    \end{tabular}
    \caption{{\bf Properties inferred from NIRCam Grism spectroscopy.} The observed redshift and fluxes of emission lines (in units of $10^{-18}$ erg/s/cm$^2$) detected in the NIRCam/WFSS spectra. The instantaneous star-formation rate (SFR) is not dust-corrected given the lack of constraints on dust in these galaxies.}
    \label{tab:ELfluxes}
\end{table}

\textbf{GN-z11} was recently observed with the NIRSpec IFU which has revealed the presence of three regions emitting either HeII (GN-z11-C) \cite{Maiolino+23_PopIII} or \Lya\ emission (GN-z11-B and a more tentative candidate) \cite{Scholtz+23}. While none of these regions show evidence of a UV continuum in deep NIRCam imaging, a tentative indication of CIV]$\lambda\lambda$1548,1551 emission in GN-z11-B and analysis of the emission in GN-z11-C \cite{Maiolino+23_PopIII} indicate they are indeed being driven by a UV-faint stellar population in these regions. As such, we conclude that GN-z11 does have spectroscopically-confirmed companion galaxies. Moreover, the presence of multiple LAEs in the local vicinity of GN-z11 suggests that peculiar velocity is not playing a vital role in the escape of \Lya\ through the neutral IGM (see \cite{Scholtz+23} for a detailed analysis of the \Lya\ emission present in this system). Moreover, the density of objects in this field makes GN-z11 a likely protocluster core \cite{Scholtz+23}.

\noindent{\bf SED fitting} 

\noindent  We then use \textsc{Bagpipes}, with the redshift fixed to that of the spectroscopic redshift of the system, using different SFHs (constant, delayed, burst and burst+constant) to fit the SED of all central and companion galaxies, as seen in Extended Figure~\ref{Efig:1} and ~\ref{Efig:1}. The best fit model is then obtained by taking the SFH that leads to the smallest Bayesian Information Criterion (BIC) as described in \cite{Laporte21}. We then report the galaxy properties returned from this method in Table~\ref{tab:Galaxy_Properties}.

\vspace{2cm}
\noindent{\bf Close companion fraction of non-LAEs}

\noindent  In order to ascertain the companion fraction of a mass-matched sample of $z>7$ galaxies, we take 30 galaxies with stellar masses from $7.4 < \rm{log(M}_{*} [$M$_{\odot}]) < 9.3$ from \cite{Laporte+22, Tang+23,Harikane+22,Bouwens+22}. We find that 47\% of these galaxies have photometric-candidate companions within 5$^{\prime \prime}$ of the central galaxy. However, many of these galaxies have not been spectroscopically followed-up and as such, we do not have constraints on their \Lya\ emission. 

Unfortunately, examples of spectroscopically confirmed $z>7$ non-\Lya\ emitting galaxies that have been observed by high-resolution spectrographs are rare in the literature. As such it is challenging to probe the companion fraction for known non-LAE galaxies. Moreover, the challenges of slit/aperture spectroscopy mean that it is incredibly rare to have spectroscopic information on the local environment of such galaxies and hence, while a galaxy may be confirmed as non-LAE, there is no information regarding the \Lya\ emission of any companion that may or may not be present. 

The recent JADES data release of NIRSpec and NIRCam observations in GOODS-S includes seven $z>7$ galaxies, observed using the NIRSpec high-resolution G140M grating, that show no \Lya\ emission \cite{Jones+23}. This does not provide any \Lya\ information on any would-be companion galaxies, however, the companion fraction for this sample, $\sim 43\%$, is once again consistent with both our larger sample, and that of Pusk\'{a}s et al. (in prep.). We therefore conclude that the $100\%$ companion fraction of our LAE sample is clearly atypical compared to the companion fractions of samples not selected for \Lya\ emission. 

\vspace{0.2cm}
\noindent{\bf Close companions are necessary but not sufficient for Ly$\alpha$ emission}

\noindent We note the recently reported $z = 9.3$ merging system \cite{Boyett+23} with an absence of detected \Lya\ emission. The reported specific-star-formation rate of this galaxy (sSFR $\sim -8$) is consistent with the high sSFR seen in our sample of merging LAEs, and therefore, one may naively posit that this is a contradiction to our results. We argue that non-detection of \Lya\ emission from a merging system is entirely reasonable and in fact our simulation modelling predicts this. While we see significant boosts in the SFR and hence intrinsic \Lya\ emission from merging systems, the escape of \Lya\ emission is only possible when the escape fraction of \Lya\ emission out of both the host galaxy and the surrounding IGM is non-zero. We explore the effect of viewing angle on the observed \Lya\ emission escaping our simulated merging systems and find that while the majority of viewing angles facilitate the escape of \Lya, several viewing angles exist for which the only \Lya\ emission escaping the galaxy is diffuse and is therefore unobservable. Moreover, without the presence of a large ionised bubble, the neutral IGM will not facilitate the escape and observation of \Lya\ emission. Ultimately we conclude that close companions are necessary but not sufficient for the observation of Ly$\alpha$ from galaxies deep in the epoch of reionisation. A combination of a large-scale ionised bubble and a preferable line-of-sight, stripped of neutral hydrogen by the interaction with the companion, are also required.

\vspace{2cm}
\noindent{\bf \large Azahar simulations}

\begin{figure}
    \includegraphics[width=0.95\columnwidth]{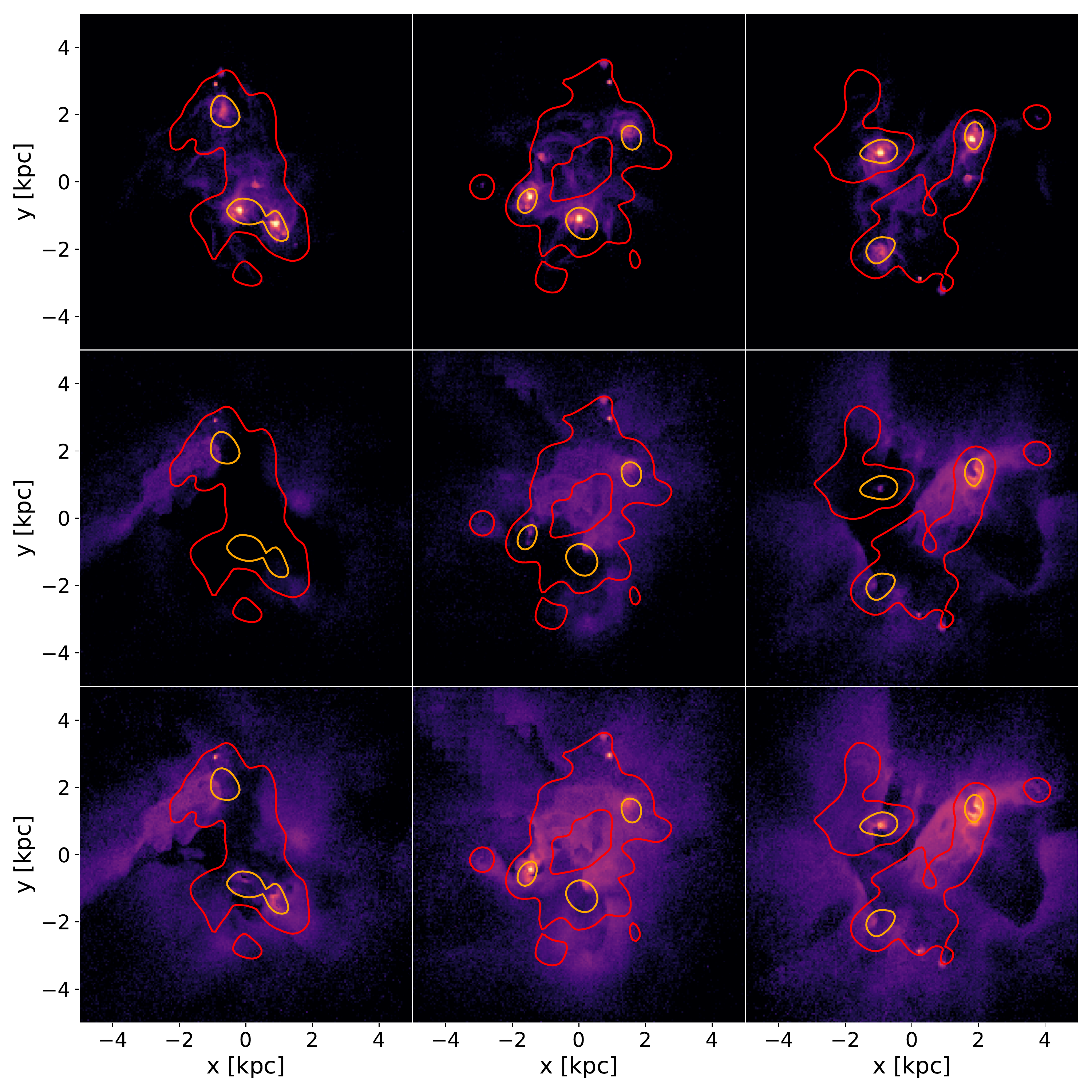}\\
    \caption{Dust opacities superimposed on synthetic \Lya\ images at $z = 7.3$. The top row shows the intrinsic \Lya\ emission, the middle row shows the scattered \Lya\ emission generated with a default dust model, and the bottom row shows the scattered \Lya\ emission  with practically no dust  (dust opacity reduced by a factor of $10^6$). The three different columns show the galaxy from three different directions. Red and orange iso-contours correspond to a dust optical depth of 0.2 at the \Lya\ wavelength and H$\alpha$ wavelength, correspondingly. Note that the dust does not act as a screen for the \Lya\ emission because of the resonant nature of \Lya\ scattering.}
    \label{Efig:6}
\end{figure}

\noindent The new \simname~simulations are a suite of multiple models spanning various combinations of canonical hydrodynamics, magnetic fields, radiative transfer, and cosmic rays physics with the aim of understanding their effects on galaxy formation as well as their complex interplay. The simulations are generated using the magneto-hydrodynamical code {\sc ramses} \cite{Teyssier2002}, which employs a constrained transport method for divergence-less evolution of the magnetic field \cite{Teyssier2006}. The main target of study for \simname~is a massive spiral galaxy with $M_\text{halo}\,(z = 1) \sim 2 \cdot 10^{12}\,M_\odot$ in a relatively large zoom-in region, approximately $8$~cMpc across along its largest axis. \simname~has a maximum spatial resolution in cells with a full cell-width of $\Delta x \sim 20\,\text{pc}$ (or equivalently, an approximate cell half-size or radius of $10\,\text{pc}$), and refinement is triggered whenever its size is larger than a quarter of its Jeans length, or it contains a total mass larger than $8\, m_\text{DM}$, where $m_\text{DM} \sim 4.5 \cdot 10^{5} M_\odot$ is the mass of dark matter particles in the zoom region. The stellar mass particle is $m_* \sim 4 \cdot 10^{4} M_\odot$. \simname~follows the set of physics presented by the pathfinder Pandora \cite{Martin-Alvarez+2023}. We provide here a brief summary, and refer the reader to that reference and Martin-Alvarez et al. (in prep.) for further details.

In this work, we investigate one of the most complete simulations in the \simname~suite, the RTnsCRiMHD model. In this manuscript, we refer to the RTnsCRiMHD simulation simply as \simname. Radiative cooling is modelled both above and below $10^4 \text{K}$ \cite{Ferland1998, Rosen1995}. The model adopts a magneto-thermo-turbulent prescription for star formation \cite{Kimm2017, Martin-Alvarez2020}, mechanical supernova (SN) feedback \cite{Kimm2014} and astrophysically-seeded magnetic fields through magnetised SN feedback \cite{Martin-Alvarez2021}. This feedback injects 1\% of the SN energy as magnetic energy, and intends to reproduce the approximate magnetisation of SN remnants. This particular simulation of \simname~also includes cosmic rays modelled as an energy density allowed to anisotropically diffuse, evolved with an implicit solver \cite{Dubois2016, Dubois2019}. In \simname, cosmic rays are exclusively sourced by SN feedback, with each event injecting 10\% of their total energy as cosmic rays. We assume a constant diffusion coefficient for cosmic rays of $\kappa_\text{CR} = 3 \cdot 10^{28}\,\text{cm}^2\,\text{s}^{-1}$, and do not account for cosmic ray streaming in this model. This model also includes on-the-fly radiative transfer \cite{Rosdahl2015}, with a configuration similar to that of the {\sc SPHINX} simulations \cite{Rosdahl2018}, featuring three energy bins for radiation spanning the ionisation energy intervals from $13.6$~eV to $24.59$~eV (HI ionisation), $24.59$~eV to $54.42$~eV (HeI ionisation), and above $54.42$~eV (HeII ionisation). Finally, recent work has shown that when expanding simulation models to account for cosmic ray feedback physics at comparable resolutions, the escape fractions of LyC photons from galaxies is lower than in their absence \cite{Farcy2022}.

Therefore, \simname~features a realistic (yet computationally taxing) ISM model that considerably approaches the resolution required to converge in the propagation and escape of ionising photons from galaxies \cite{Kimm2014}. We note that, while close, our resolution still falls short from that regime and that even more sophisticated ISM models \cite{Katz2022} may be important for a complete understanding of photon propagation in the ISM. Opportunely, a considerable part of our results relies on gas being ejected from galaxies during merger events, which implies that our estimate of LAE detectability during these events will be more resilient to these caveats.

\vspace{0.2cm}
\noindent \textbf{Simulated ionised bubble evolution in the neutral IGM}

\noindent In the high resolution patch of the Universe probed by our simulation ($\lesssim 8$~Mpc on a side), centred on the main progenitor of our $z \sim 1$ galaxy, the first ionised bubbles emerge at $z \sim 20$. By $z \sim 15$ several ionised bubbles exist that are a few (physical) kpc in radius, and rapidly begin merging as redshift approaches $z \sim 10$. This drastically increases the mean free path of ionising photons, with the coalesced bubble reaching scales of $100$~kpc. The resulting main bubble continues to grow, reaching a radius $\sim 0.5$~Mpc at $z \sim 7.3$. 

For intrinsically bright LAEs, as studied here, a $\lesssim 0.5$~Mpc local ionised bubble may be sufficient for these sources to be detectable \cite{Hayes+23}. We further note that due to ongoing mergers our simulated galaxies have considerable relative peculiar velocities of up to $230$~km~s$^{-1}$. Sufficiently large velocities relative to the intervening neutral IGM gas may reduce damping wing suppression for particular lines-of-sight, hence requiring smaller local ionised bubbles to facilitate the detection of \Lya. Our simulation also does not consider star formation outside of the high-resolution zoom-in region. Hence, our simulated ionised bubbles could be even larger in size, e.g. due to the presence of a neighbouring, large-scale overdensity or large-scale clustering \cite{Weinberger2019} on spatial scales beyond our zoom-in region.

Nevertheless, knowing from our observations that the LAEs sit in sufficiently large ionised bubbles, here we primarily focus on the probability of \Lya\ photons escaping the host galaxy and its intermediate surroundings. We find that a large amount of gas remains neutral within the local ionised bubble, actively feeding the galaxies along the cosmic web. 

\vspace{0.2cm}
\noindent \textbf{Galaxy selection and measurements in the simulations}

\noindent In order to select galaxies in our simulation, we employ the {\sc halomaker} software \cite{Tweed2009} to detect and characterise dark matter halos. We identify the three progenitor galaxies of interest (as well as the galaxy merging with the main progenitor at $z \sim 8.1$) and follow them through time by tracking their innermost stellar particles. Their centres are determined using a shrinking spheres algorithm applied to their stellar component \cite{Power2003}, and they are assigned their corresponding halo as obtained by {\sc halomaker}. In order to select the system for study, we broadly reviewed the evolution of all galaxies with stellar masses $M_* > 10^{7} M_\odot$ in the redshift interval $z \in [9.0, 6.0]$. The most promising candidate, the main progenitor of the main \simname~galaxy was selected for further investigation. 

In order to assign measurements to individual galaxies, we measure values within their galactic region, defined by the radius $r_\text{gal} < 0.2\; r_\text{rvir}$. During the merging stages, we employ $r = \text{min}[r_\text{gal}, 0.45 D_{ij}]$, where $D_{ij}$ is the distance between the $i$ and $j$ progenitors, down to a minimum distance of $1.5$~kpc. 

We have briefly reviewed the observability of lower mass pairs within the simulated domain. While these present a similar behaviour during mergers to that of our main studied system, their observability and that of other galaxies with similar masses remains low. This is due to their low stellar masses generating low luminosities. Their low stellar masses also make them more vulnerable to disruption during powerful star formation bursts, which increases their escape fractions \cite{Rosdahl2022}. Despite this, most isolated systems present quiet star formation histories and in the absence of mergers, have low observabilities throughout their evolution.

\noindent{\bf RASCAS post-processing}

\noindent We post-process the simulation with the publicly available, massively parallel code RASCAS \cite{Michel-Dansac20}, for modelling the \Lya\ emission and resonant scattering in our simulations. RASCAS accounts for two sources for the \Lya\ emission, recombination and collisional excitation. For recombination, RASCAS adopts the case B recombination coefficient from \cite{Hui&Gnedin97}, and the fraction of recombinations producing \Lya\ photons from \cite{Cantalupo08}. For collisional excitation, RASCAS uses the fitting function for the collisional excitation rate from level 1s to 2p from \cite{Goerdt10}. We cast $N_{\rm MC} = 10^6$ Monte Carlo photon packets to sample the real \Lya\ photon distributions, from recombination and collisional excitation, respectively. We sample $N_{\rm MC} = 10^7$ photon packets for the images along the LOS in Extended Figure~\ref{Efig:3}.  

After each scattering event, the \Lya\ photon changes its frequency and direction according to a phase function as implemented in RASCAS. RASCAS adopts the phase function in \cite{Hamilton40, Dijkstra&Loeb08} for the scattering of \Lya\ photons around line centre and Rayleigh scattering for \Lya\ photons in the line wing. At high H~\textsc{i} column density, \Lya\ photons will scatter many times locally until they shift in frequency large enough so that they can have a long mean free path. To reduce the associated computation cost, RASCAS adopts a core-skipping mechanism \cite{Smith15} to transit the photons to line wings while avoiding local scattering in space. RASCAS also implements the recoil effect and the transition due to deuterium with an abundance of D/H = $3 \times 10^{-5}$. The dust distribution is modelled according to the dust model described in \cite{Laursen09b}. Dust can either scatter or absorb \Lya\ photons. The probability of scattering is given by the dust albedo $a_{\rm dust} = 0.32$ following \cite{Li&Draine01}. The dust scattering with the Henyey-Greenstein phase function \cite{Henyey&Greenstein41} and the asymmetry parameter is set to $g=0.73$ following \cite{Li&Draine01}. We generate synthetic images and spectra with a peeling algorithm described in \cite{Yusef-Zadeh84, Wood&Reynold99, Costa22}, along 12 lines-of-sight uniformly sampled using the healpix algorithm \cite{Gorski05}.

Finally, we note that the dust distribution has a large effect on the \Lya\ radiative transfer as well as the \OIII and \Hb emission that is used to infer the intrinsic \Lya\ emission from observations. In Extended Figure~\ref{Efig:6}, we show how reducing the amount of dust changes the observed \Lya\ images. We see the dust is able to completely mask the \Lya\ emission of some individual galaxies as resonant scattering of H~\textsc{i} increases the probability of \Lya\ photons encountering dust grains. The dust can suppress the \Ha emission by up to 20\% around the center of the three merging galaxies, if we assume intrinsic \Lya\ can be converted directly to intrinsic \Ha emission. We therefore conclude that inclusion of dust modelling, as done here, is fundamental for understanding observed \Lya\ emission. However, our results regarding the boosted intrinsic \Lya\ emission in star formation bursts and enhanced \Lya\ escape fraction in mergers, due to the channels cleared of local neutral hydrogen are robust to the assumed dust model.

We further consider the \Lya\ profiles escaping along the line-of-sight shown on the upper panel of Extended Figure~\ref{Efig:3} at $z = 7.3$ after applying a reasonable IGM damping wing from \cite{Keating23}. We note that the shape of the \Lya\ profile is very complex and sensitive to both the LOS \cite{Blaizot+23} and IGM attenuation \cite{Witten+23}. As such, further analysis of this is needed to constrain the expected \Lya\ profiles from simulations requiring a larger simulated galaxy sample with careful IGM attenuation modelling as well as a larger observational sample to compare to, which is beyond the scope of this work. However, as an immediate consistency test, we take the \Lya\ profile from our simulated system at $z = 7.3$ after IGM attenuation \cite{Keating23}  and compare the velocity offset and FWHM to that reported in our observational comparison, EGSY8p68. We find that the observed \Lya\ profile \cite{Larson23}  is well matched with the simulated profile's velocity offset of $\sim 200$ km/s and FWHM of $\sim 200$ km/s, and shows a clear asymmetric line shape as observed.

\newpage

 \noindent{\bf \large Data availability}

\noindent The JWST data used in this analysis is publicly available from the STScI MAST Archive.

\vspace{0.2cm}
\noindent{\bf \large Code availability}

\noindent The \textsc{WebbPSF} tool can be found at \url{https://www.stsci.edu/jwst/science-planning/proposal-planning-toolbox/psf-simulation-tool}. The standard MOSFIRE data reduction pipeline can be found at \url{https://keck-datareductionpipelines.github.io/MosfireDRP/} and EsoReflex can be found at \url{https://www.eso.org/sci/software/esoreflex/}.

\vspace{0.2cm}
\noindent{\bf \large Acknowledgments}

\noindent The authors would like to acknowledge the work of the PRIMER core team in obtaining and reducing the NIRCam data for the COSMOS field used in this work (PI: Dunlop, ID: 1837). The authors would also like to acknowledge the work of the CEERS team in obtaining and reducing the NIRCam data of the EGS field used in this work (PI: Finkelstein, ID: 1345). The authors would like to acknowledge the FRESCO team's work in obtaining observations of the GOODS-S and GOODS-N fields (PI: Oesch, ID: 1895). The authors would also like to thank Joki Rosdahl for their support with the generation of radiative transfer RAMSES simulations, and Thibault Garel for their support with the usage of RASCAS. The authors also thank Mengtao Tang, Daniel Stark and Hannah \"Ubler for providing additional information on the properties of some of our sample. 

Support for this work was provided by NASA through grant JWST-GO-01837 awarded by the Space Telescope Science Institute, which is operated by the Association of Universities for Research in Astronomy, Inc., under NASA contract NAS 5-26555. This research has made use of the Keck Observatory Archive (KOA), which is operated by the W. M. Keck Observatory and the NASA Exoplanet Science Institute (NExScI), under contract with the National Aeronautics and Space Administration. Based on observations collected at the European Southern Observatory. This work used the DiRAC@Durham facility managed by the Institute for Computational Cosmology on behalf of the STFC DiRAC HPC Facility (www.dirac.ac.uk). The equipment was funded by BEIS capital funding via STFC capital grants ST/P002293/1, ST/R002371/1 and ST/S002502/1, Durham University and STFC operations grant ST/R000832/1. DiRAC is part of the National e-Infrastructure. CW thanks the Science and Technology Facilities Council (STFC) for a PhD studentship, funded by UKRI grant 2602262. NL acknowledges support from the Kavli foundation. DS, MGH and JSD acknowledge STFC support. PS acknowledges INAF Mini Grant 2022 “The evolution of passive galaxies through cosmic time”. RSE acknowledges financial support from ERC Advanced Grant FP7/669253. PGP-G acknowledges support  from  Spanish  Ministerio  de  Ciencia e Innovaci\'on MCIN/AEI/10.13039/501100011033 through grant PGC2018-093499-B-I00. DP acknowledges support by the Huo Family Foundation through a P.C. Ho PhD Studentship. RM, WB and WM acknowledge support by the Science and Technology Facilities Council (STFC) and ERC Advanced Grant 695671 "QUENCH". RM also acknowledges funding from a research professorship from the Royal Society. This research was supported in part by the National Science Foundation under Grant No. NSF PHY-1748958.

\vspace{0.2cm}
 \noindent{\bf \large Author Contributions} 
 
 \noindent CW reduced and analysed NIRCam imaging and WFSS, MOSFIRE and X-shooter data. NL extracted the photometry of each candidate and performed the SED-fitting analysis. SMA developed the \textit{Azahar} simulations, performed their basic analysis and extracted the Lyman-$\alpha$ luminosity for all the merging systems. YY ran the RASCAS simulations and showed the influence of line of sights. DS and MGH coordinated the simulations part of this paper. CW, NL, SMA, DS, YY and MGH wrote the paper, and developed the main interpretation of the results. RM, GRB, RSE, WM, WB, DP, CS and HK were key in interpreting the observational results. JSD, RSE, NG, GI, AMK, DM, PGP-G and PS were involved in obtaining and reducing the data for the PRIMER program which was used in this study. All authors discussed the results and commented on the manuscript.

\vspace{0.2cm}
 \noindent{\bf \large Competing interests}

\noindent The authors declare no competing interests.

\begin{appendices}

\end{appendices}

\bibliography{sn-bibliography}

\end{document}